\documentclass[prb, preprint,showpacs]{revtex4}
\usepackage{epsfig,multirow}

\begin{document}
\title{Few-electron artificial molecules formed by laterally coupled quantum rings}
\author{T. Chwiej} \affiliation{Faculty of Physics and Applied
Computer Science, AGH University of Science and Technology,\\ al.
Mickiewicza 30, 30-059 Krak\'ow, Poland}
\author{B. Szafran}
\affiliation{Faculty of Physics and Applied Computer Science, AGH
University of Science and Technology,\\ al. Mickiewicza 30, 30-059
Krak\'ow, Poland}

\begin{abstract}
We study the artificial molecular states formed in laterally coupled double
semiconductor nanorings by systems containing one, two and three electrons. An interplay of
the interring tunneling and the electron-electron interaction is described and its consequences
for the magnetization and charging properties of the system are determined.
It is shown that
both the magnetic dipole moment generated by the double ring structure and the chemical potential of the system
as function of the external magnetic field strongly depend on the number of electrons and the interring barrier
thickness. Both the magnetization and chemical potentials exhibit cusps at the magnetic fields inducing
ground-state
parity and / or spin transformations. The symmetry transformations are discussed for various tunnel coupling strengths:
from rings coupled only electrostatically to the limit of coalesced rings.
We find that in the ground-states for rings of different radii the magnetic field transfers
the electron charge from one ring to the other. The calculations
are performed with the configuration interaction method based on an approach of Gaussian
functions centered on a rectangular array of points covering the studied structure.
Electron-electron correlation is also discussed.
\end{abstract}
\pacs{73.40.Gk} \maketitle
\section{Introduction}
In the eighties semiconductor rings of micrometer size were investigated in quest
for signatures of the Aharonov-Bohm effect \cite{ab} in conductance measurements.\cite{buti,timp} In the next decade
magnetization produced by persistent currents circulating around semiconductor rings
was measured.\cite{ms} Subsequent technological advances allowed for fabrication of rings with nanometer radii
with detectable quantum size effects. Nowadays, the quantum rings are produced with
the etching \cite{hawrylak} or surface oxidation techniques \cite{so} as well as grown by self-assembly.\cite{lorke}
Transport experiments are performed on open quantum rings,\cite{o} while the closed rings
are studied in the context of single-electron charging \cite{lorke} or optical properties.\cite{gz,jask,puente} Recently, magnetization
signal of large ensembles of semiconductor self-assembled nanorings has been detected. \cite{fomin}

The theoretical literature on quantum rings is very rich. The authors mainly concentrated on properties
of a single isolated quantum ring.\cite{jask,puente,rev,pcct0,pcct1,pcct2,pcct3,pcct4,foomin}
At present there is a growing interest in systems of multiple quantum rings including arrays of quantum rings.\cite{fo,moje}
Moreover, double rings are produced in both concentric \cite{dcqr1,dcqr2,dcqr3} and vertical\cite{vqr} configurations.
The tunnel and electrostatic coupling was theoretically studied for both
concentric \cite{tst,cl1,cl2,pi,cast,zc1,zc2,zc4} and vertically stacked rings.\cite{vertwo}
Recently, states of a single-electron in a pair of laterally coupled quantum rings
were described.\cite{ostatni} The magnetization generated by  planar arrays of interacting quantum rings \cite{moje}
for neglected tunneling between the rings was also discussed.
The purpose of the present paper is to describe
the system of up to three electrons in an artificial molecule
formed by two quantum rings with the account taken for both the tunnel coupling
and the electron-electron interaction. We investigate the competition between the tunnel and Coulomb coupling,
the electron-electron correlation as well as the charging and magnetization properties. The numerical results
are provided for the etched InGaAs/GaAs rings \cite{hawrylak} with low indium concentration and consequently low
potential depth which favors  electron tunneling
between the rings.

The evolution of the single-electron ground-state with the magnetic field for artificial molecules formed by quantum rings
is significantly more complex than for the double quantum dots. In double dots the role of the magnetic field
for the eigenstates of the single electron is limited to reduction of the interdot tunnel coupling.\cite{rcc} In double rings
the magnetic field drives the angular momentum transitions of the single-electron within each of the rings.\cite{rev}  When the tunnel coupling
is activated the single-electron ground-state becomes localized at the contact between the rings \cite{ostatni} with the strength
of the localization oscillating in function of the magnetic field with a period corresponding to the flux quantum threading
the rings. For identical rings the single-electron ground state corresponds to a binding orbital\cite{ostatni} and possesses an even spatial parity
irrespective of the value of the magnetic field.
In this paper we show that both the spin and spatial symmetry transitions occur with the external magnetic field\cite{mani}
double rings containing few electrons. We find that the symmetry transformations in the external magnetic field depend strongly
on the interring barrier thickness and they have quite a different character for various numbers of electrons confined
within the double ring structure. Since the transformations influence strongly the charging and magnetic
properties, an evidence of the tunnel coupling between the rings should be detectable by measurements of chemical potentials and magnetization.

Our discussion covers asymmetric configurations composed of rings of different radii for which
we find that the magnetic field induces an oscillatory switching of the ground-state localization from one ring to the other.
This effect can be used to transfer the electron between the rings without a need for applying an external electric field to the system.

In this work we use the configuration interaction approach which allows for a numerically exact solution of the Schroedinger equation
for a few electrons. Application of the configuration interaction method to the laterally coupled double ring structure in the external magnetic structure is challenging
as compared to both laterally coupled dots and double rings in concentric and vertical configurations.
The basis used for the configuration interaction calculation has to keep track of the angular momentum transitions which occur in
each of the rings separately but the total angular momentum cannot be used as a quantum number for selection of the basis set since the system
does not possess circular symmetry unlike concentric double or vertically stacked rings. For the purpose of the present study we developed
quite a powerful technique in which a Gaussian functions are used with centers distributed on a regular array.
The presented technique is universal and can be applied to few-electron systems
in arbitrary smooth confinement potentials.

This paper is organized as follows. In Section II we present the model Hamiltonian, the confinement potential
and the configuration interaction approach based on a mesh of Gaussian functions. A test of the method for two-electron states
is presented and the limitations of the approach are explained on this example. The results are given in Section III.
We start by single-electron states of a single and identical double rings. The discussion is then extended to few-electron states and non-identical rings. The summary and conclusions are given in Section IV.

\section{Theory}

We consider the following few-electron Hamiltonian
\begin{equation}
\widehat{H}=\sum_{i=1}^{N}\widehat{h}_{i}+\sum_{i=1,j>i}^{N}\frac{1}{\varepsilon r_{ij}}
\label{hn}
\end{equation}
where $\widehat{h}_{i}$ is single electron energy operator and apply the configuration
interaction approach in which $\widehat{H}$ operator is diagonalized in a basis of
many electron wavefunctions with determined values of total spin S and its projection on z axis
$S_{z}$.  The basis functions are generated with the help of projection operator \cite{low}
as the linear combinations of Slater
determinants built of eigenfunctions of a single-electron Hamiltonian:
\begin{equation}
\widehat{h}_{i}=\frac{(\widehat{\bf{p}}+e{\bf{A}}({\bf{r}}_{i}))^{2}}{2m^{*}}+V_{ext}
(\bf { r } _ { i } ).
\label{hi}
\end{equation}
We use the vector potential in a symmetric gauge: ${\bf {A}}({\bf{r}})=B/2(-y,x,0)$.
The single-electron eigenproblem is diagonalized in a basis
\begin{equation}
\phi_{i}({\bf{r}})=\sum_{\alpha=1}^{N}C_{\alpha}^{(i)}f_{\alpha}
\label{jedn},
\end{equation}
where the basis function have the form
\begin{eqnarray}
f_{\alpha}({\bf{r}})=\exp\bigg(-\frac{({\bf {r}}-{\bf{R}}_{\alpha})^2} {2\sigma^{2} } \bigg)
\exp\bigg(-\frac{ie}{2\hbar} ({\bf {B}}\times {\bf{R}}_{\alpha})\cdot {\bf {r}}
\bigg).
\label{fa}
\end{eqnarray}
In Eq. \ref{fa} the probability density associated with each basis function is a Gaussian centered at point ${\bf {R}}_{\alpha}=(x_{\alpha},y_{\alpha})$, $\sigma$  is responsible for the strength of the localization and the term with the imaginary exponent introduces the magnetic translation,
which ensured the gauge invariance of the basis, or in other word equivalence of all the basis functions
irrespective of the localization center.
The centers are distributed on a regular mesh of points (see below).
The matrix elements of the Coulomb interaction are integrated according to the procedure explained in the Appendix.

We model the potential of a single two-dimensional quantum ring by the formula:
\begin{equation}
V_{l(r)}(\vec{r})=-V_{0}\exp\bigg (-\bigg | \frac{|\vec{r}-\vec{R}_{l(r)}|-R_{0}}{\sigma_{0}}
\bigg |^\alpha \bigg )
\end{equation}
where $\vec{R}_{l(r)}$ denotes the center of the left (right) ring.
We assume $\alpha=20$ for which the potential is nearly a square quantum well.
The other parameters of the rings are adopted for the etched In$_{0.1}$Ga$_{0.9}$As/GaAs quantum rings \cite{hawrylak,ostatni}:
$V_{0}=50$ meV, the radius of single quantum ring $R_{0}=30$ nm, $\sigma_{0}=20$ nm, effective mass of an electron
 $m^{*}=0.05$ and dielectric constant $\varepsilon=12.4$.

\begin{figure*}[ht!]
\centerline{\hbox{\epsfysize=70mm
               \epsfbox {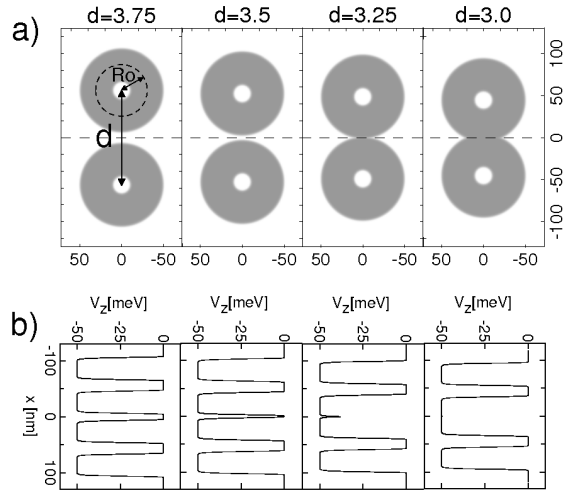}}}
\caption{Double ring structures for various parameters of the interring distance $d$ and the confinement potential cross
sections through the centers of the rings (b) for three values of parameter d.}
\label{pot}
\end{figure*}

Confinement potential of the double ring structure is assumed in form,
\begin{equation}
V_{ext}({\bf{r}})=\min\bigg(V_{l}({\bf{r}}),V_{r}({\bf{r}}) \bigg) \label{poteec}
\end{equation}
where $V_{l}({\bf{r}}),V_{r}({\bf{r}})$ are the confinement potentials of the left and the right
quantum ring respectively.  We use a dimensionless parameter $d$ which describes the distance between
the centers of two coupled rings:
\begin{equation}
d=\frac{|\vec{R}_{l}-\vec{R}_{r}|}{R_{0}}
\end{equation}
The confinement potentials of two laterally coupled quantum rings
for various $d$ are shown in Fig. \ref{pot}.\\
The centers of the single electron basis functions (\ref{fa}) are
distributed on a mesh of $k_x\times k_y$ points, where  $k_{x}=\frac{(4.8+d)}{ 4.8}\times k_{y}$, $k_{y}=40$.
For all functions $f_{\alpha}$ value of  $\sigma$  parameter is identical and equals
$\sigma=\frac{4.8R_{0}}{1.4k_{y}}$. This value was optimized for the description of
the low-energy part of the single-electron spectrum.

\begin{figure*}[ht!]
\centerline{\hbox{\epsfysize=50mm
               \epsfbox {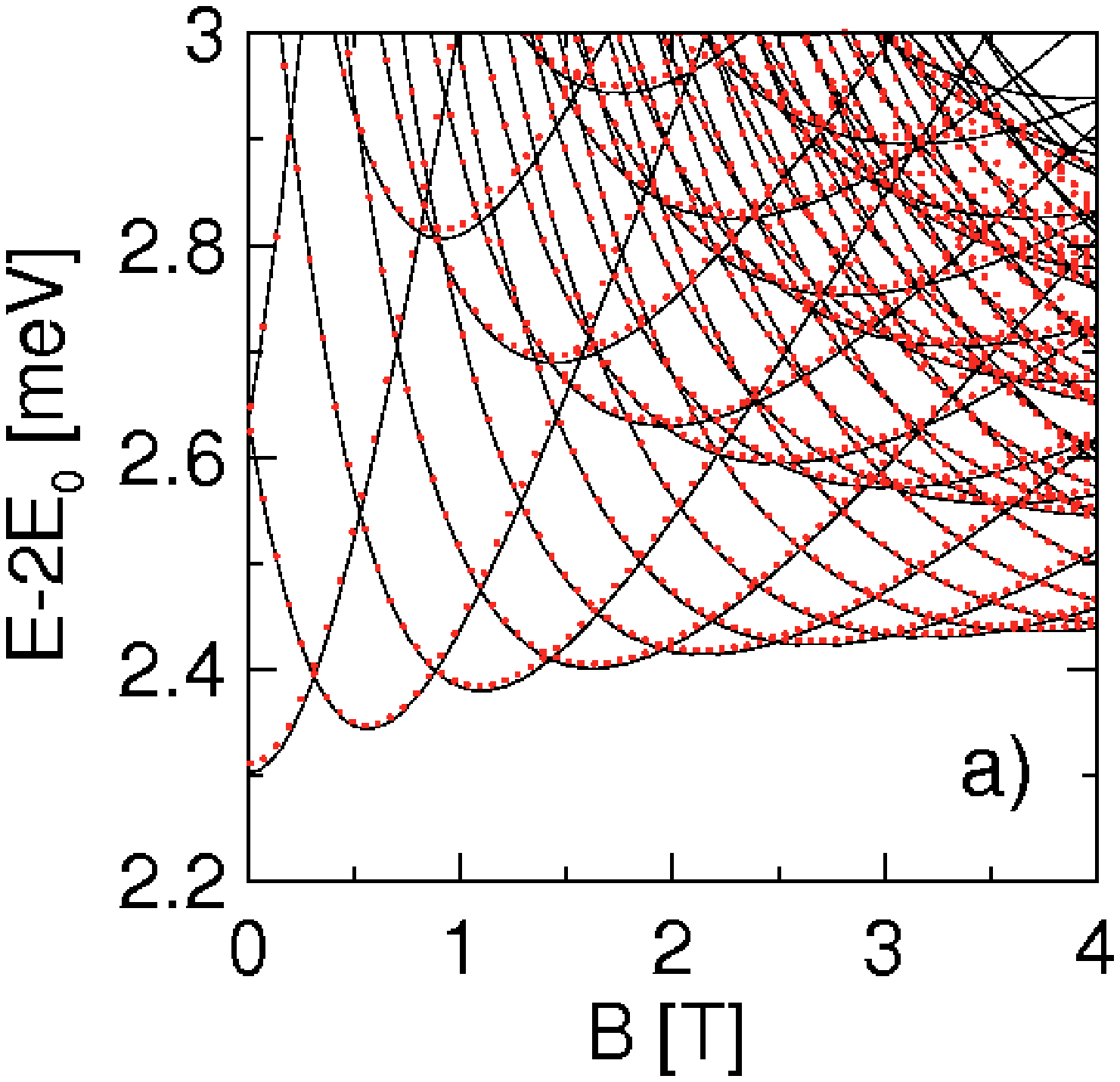}\epsfysize=50mm
               \epsfbox {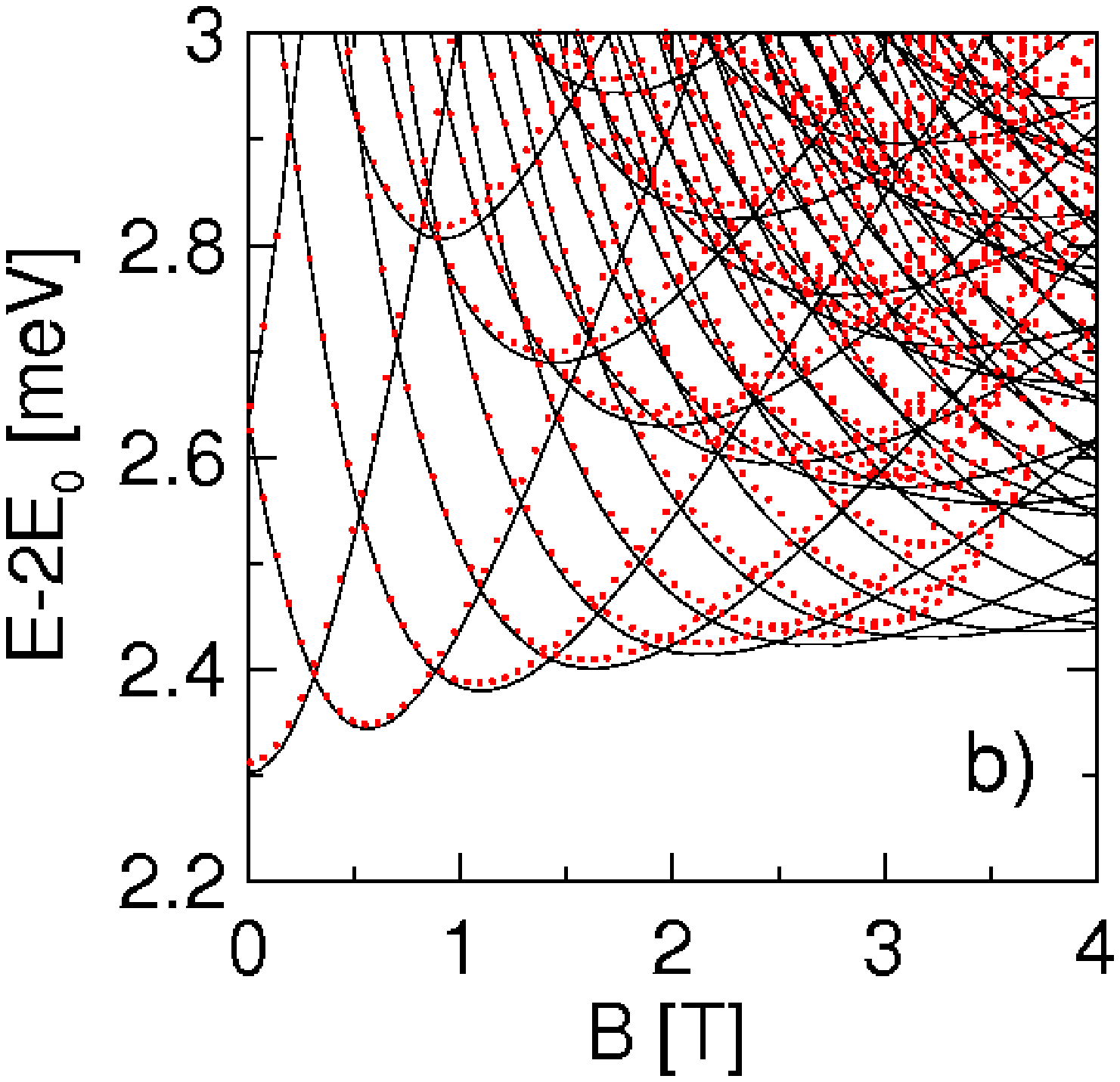}}}
\caption{Comparison of the exact two-electron energy spectrum (solid lines) for the 2D harmonic
oscillator confinement potential and the results obtained in the present CI method (red dots) for
the mesh of $50\times 50$ and $20\times 20$ Gaussians.
The energy spectra are calculated with respect to the non-interacting ground state
energy equal to double of the Fock-Darwin\cite{mani} ground-state energy level.} \label{testen}
\end{figure*}

We have tested our method for the problem of two electrons confined in two-dimensional harmonic
oscillator potential,  which can be easily solved with the center-of-mass separation technique with
an arbitrary precision (the results
can be treated as ''exact'').
We assumed the oscillator energy
$\hbar \omega=1$ meV and the calculations were performed in the square region of size $20l_{o}\times 20l_{o}$
where $l_{0}=\sqrt{\hbar/m^{*}\omega}$ is the oscillator length.
We have solved the eigenproblem with the configuration interaction method for basis containing  $N=20\times 20$ and  $N=50\times 50$ centers.
The exact energy spectra and the results of the present approach are plotted in function of $B$
in Fig. \ref{testen}.

For the basis with $N=50\times 50 $ elements, the lower part of the energy spectrum agrees very well
with the exact results for magnetic fields up to 4 T. However, for basis with $N=20\times 20$ elements, the
energy of the ground state is visibly overestimated and the overestimate becomes quite large above 3 T.

\begin{figure*}[ht!]
\centerline{\hbox{\epsfysize=50mm
               \epsfbox{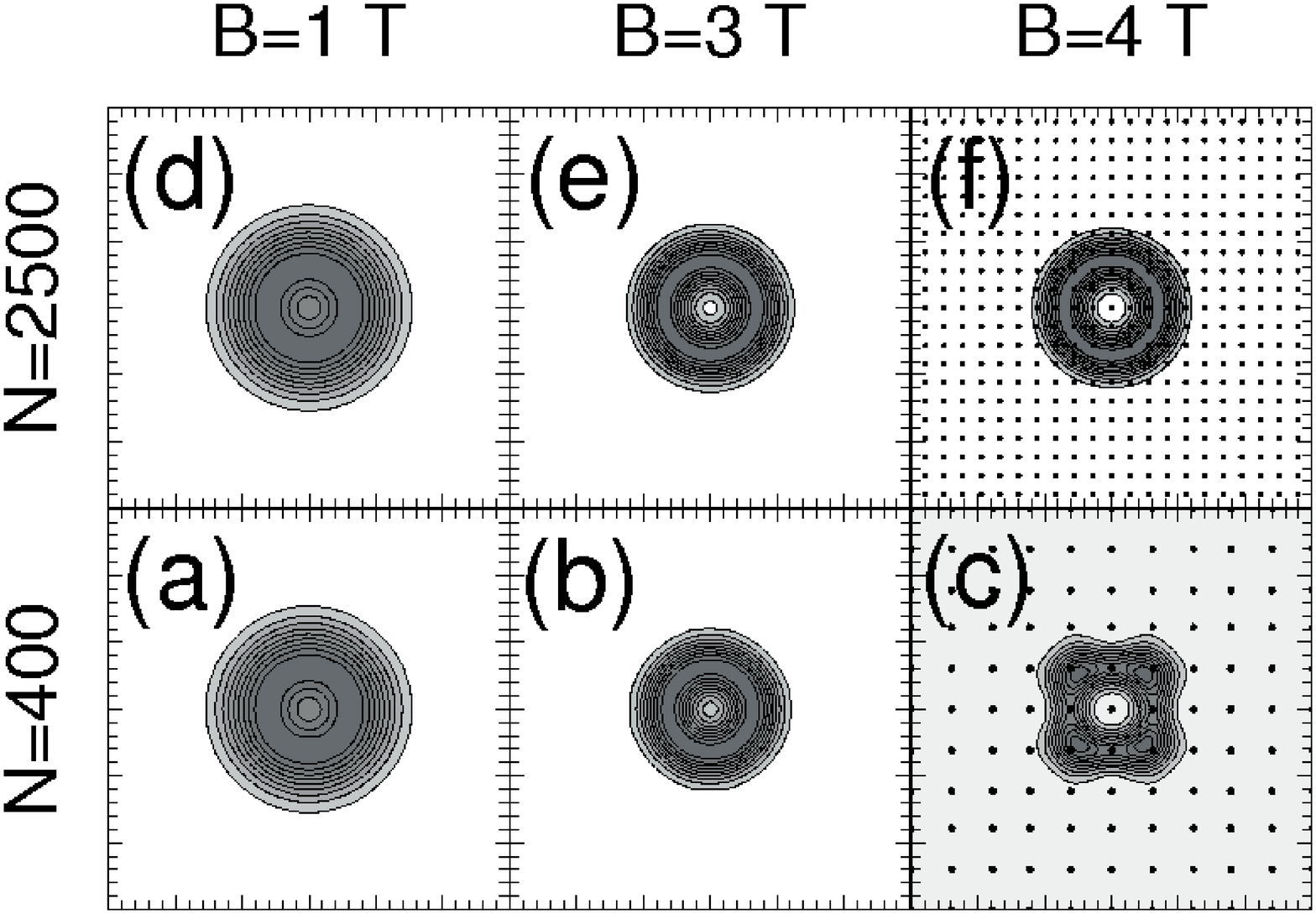}\epsfysize=50mm
               \epsfbox{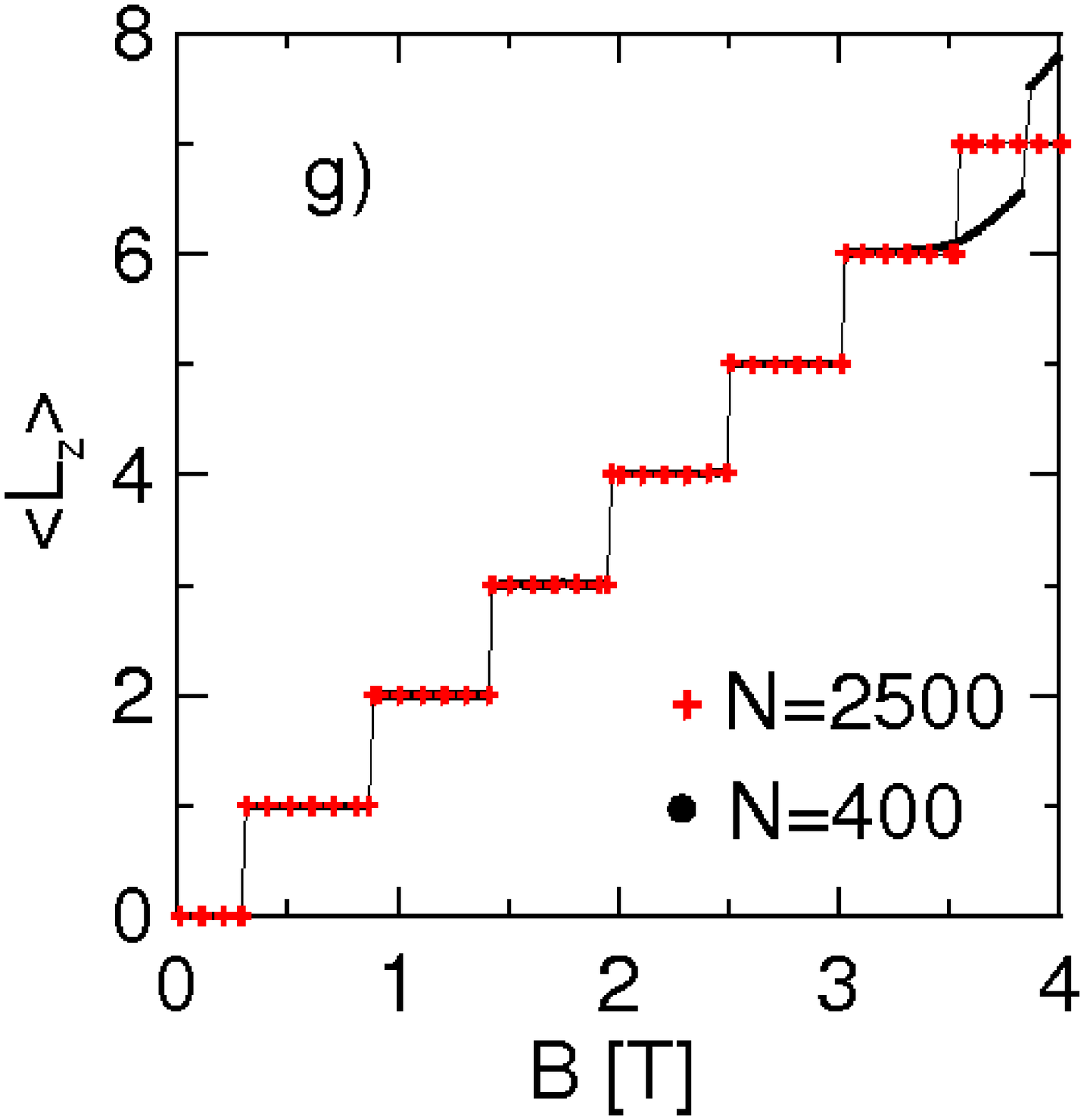}}}
\caption{(a-f) The total probability densities for $N=50\times50$ and $N=20\times 20$ basis centers
for two interacting electrons confined in a two-dimensional harmonic oscillator potential ($\hbar\omega=1$ meV) in the
singlet states. The dots on (c) and (f) indicate the centers of Gaussian functions. (g)
Expectation values the total angular momentum for $20\times 20$ elements (points)
and $50\times 50$ Gaussian (crosses).}
\label{gesttest}
\end{figure*}

For both basis the probability densities are circularly symmetric about the center of the dot for
$B=1$ T and $B=3$ T (see Fig. \ref{gesttest}(a,b) and \ref{gesttest}(d,e)).
For $B=4$ T only the probability density calculated for $N=50\times 50$ elements still preserves the confinement
potential symmetry. In the right panel of Fig. \ref{gesttest} we plot the mean value of the angular momentum
for the smaller basis. With the magnetic field the ground-state becomes more strongly localized and
acquires a high angular momentum. For fixed number of centers there is a limitation
to the maximal angular momentum that can be described in the single-electron basis.  The single-electron states of
high angular momenta correspond to sign oscillations of  both the real and imaginary part over the angular variable.
Clearly, the number of centers in the region where the probability density is non-zero sets the limitation
to the maximal frequency of the oscillations. For $B=4$ T the angular momentum obtained in the basis of $20\times 20$ elements
is non-integer  $\langle L_{z}\rangle=6.86$ and smaller than the exact ground-state value of $\langle L_{z}\rangle=8$,
which is nevertheless reproduced by the $50\times 50$ basis. Concluding, the present approach allows for a
nearly exact solution of the few-electron Schroedinger equation. There is a limitation to the maximal
angular momentum that can be accounted for, but the accuracy of the results is easily verified by comparing results
of the method produced by various meshes.


\section{Results}
\subsection{Single-electron single-ring states}
\begin{figure}[ht!]
\centerline{\hbox{\epsfysize=50mm
               \epsfbox {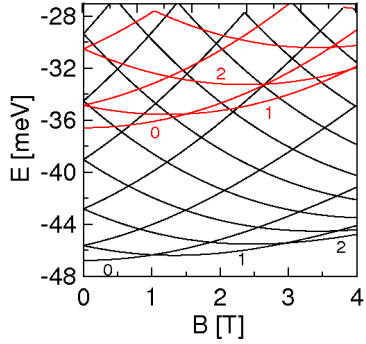}\hfill}}
\caption{Single-electron spectra for a single quantum ring. Black
curves show the energy levels associated with the lowest radial
state, and the red curves the energy levels related to the first
radial excitation. Numbers in the figure indicate the angular
momentum. \label{1e1p}}
\end{figure}

The single-electron spectrum for a single quantum ring with
potential (\ref{poteec}) is shown in Fig. \ref{1e1p}. The energy levels
plotted with black lines correspond to the lowest state of the radial
quantization and the red ones to the first radial excitation. The
energy spacing between these two branches is only about 10 meV,
which results of the small depth of the studied structure.

\begin{figure}[ht!]
\centerline{\hbox{\epsfysize=50mm
               \epsfbox {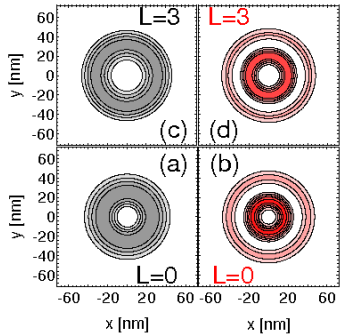}\hfill}}
\caption{Electron density for the single electron in a single ring
in zero magnetic field. Plots (a,b) correspond to zero angular
momentum and (c,d) to $L=3$. Densities in (a,c) are obtained for the
lowest-energy radial state, and (c,d) for the first radial
excitation. \label{rojejr}}
\end{figure}

 Numbers in Figure \ref{1e1p} denote the angular
momentum. In the ground-state the magnetic ``period" of the
ground-state angular momentum transition is $B_p=1.05$ T which corresponds
to the flux quantum ($B_p\pi R^2=e/h$) for a strictly one-dimensional
ring of radius $R_{1D}=35.4$ nm, slightly larger than the $R_0$ parameter (assumed equal to $30$ nm),
and closer to the average distance from the center of the ring 37 nm calculated
over the radial coordinate of the ring.  The
magnetic period for the first radial excitation is much larger and
equals $B_p=1.5$ T, which corresponds to the 1D ring of $R_{1D}=28.9$ nm.
The smaller effective radius for the excited radial state than for
the ground-state may be surprising since usually the excited
states occupy larger area than the ground-state. In order to explain
this feature we plotted the electron densities in Fig. \ref{rojejr}
for zero magnetic field. Fig. \ref{rojejr}(a) shows the density for
the ground-state and Fig. \ref{rojejr}(b) the density for the $L=0$
state of the first radial excitation. Fig. \ref{rojejr} (c) and (d) show the densities
of the angular momentum $L=3$ eigenstates -- the lowest-energy state and the first excited state, respectively.
 We notice that
the radial wave functions for $L=0$ are more strongly localized than the ones for
 $L=3$. This is due to the centrifugal effective potential $\frac{\hbar^2L^2}{2m\rho^2}$
 present in the   single-electron Hamiltonian for a single ring written in cylindrical coordinates
 for the angular momentum $L$ eigenstate
 \begin{equation}
 H=-\frac{\hbar^2\partial^2}{2m\partial \rho^2}-\frac{\hbar^2\partial}{2m\rho \partial \rho}+\frac{\hbar^2L^2}{2m\rho^2}+V(\rho)+\frac{m\omega_c^2}{8}\rho^2
 -\frac{1}{2}\hbar\omega_cL,\label{Eq1}
 \end{equation}
 where $\rho=\sqrt{x^2+y^2}$.
Fig. \ref{rojejr} shows that states that correspond to the radial
excitation with electron density forming two concentric rings occupy indeed a larger area than those corresponding to the lowest radial state.
However, the inner density ring contains
most of the electron charge and its radius is smaller then the electron density ring
in the lowest-energy radial state [cf. Fig. \ref{rojejr}(a) and (b), as well as Fig. \ref{rojejr}(c) and (d)],  which explains why the effective radius
value $R_{1D}$ obtained for the branch of the excited energy levels
is smaller than for the ground state.

\subsection{Single electron states for a pair of identical rings}
\begin{figure*}[ht!]
\centerline{\hbox{\epsfysize=50mm
               \epsfbox {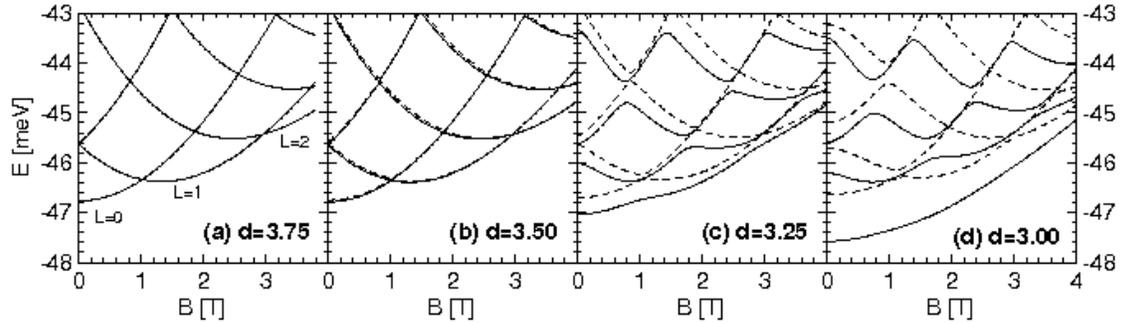}\hfill}}
\caption{Single-electron spectra for a pair of identical rings for
various interring distances. Solid lines show the even parity energy
levels and the dashed lines the odd parity energy levels. In (a) the
angular momenta with respect to the center of the ring are listed.
\label{1eip}}
\end{figure*}

The coupling between the rings is of a pure tunnel character for the
single-electron states. The single-electron energy spectra are given
in Fig. \ref{1eip} for various interring distances. For $d=3.75$ the
tunnel coupling is negligible and the spectrum [see Fig.
\ref{1eip}(a)] consists of energy levels which are twofold
degenerate with respect to the parity. This energy spectrum is in
fact identical with the one of a single separate ring (see Fig.
\ref{1e1p}). In Fig. \ref{1eip}(b) plotted for $d=3.5$, we spot
lifting of the even-odd degeneracy which occurs when the interring
tunnel coupling is activated. Note that when the degeneracy with
respect to the parity is lifted, lower energy level always
corresponds to the even parity energy level, i.e. to the binding
orbital. The even-odd degeneracy lifted by the tunnel coupling is
restored for higher magnetic fields. This results of the attenuation
of the tunnel coupling by the magnetic field which enhances electron
localization within each of the rings. For $d=3.25$ -- when
the rings nearly touch one another -- the even and odd energy levels
differ significantly in the energy. We observe a pronounced avoided
crossings in the spectrum that occur separately for the even and odd
energy levels. The pattern of the avoided crossings in the even and
odd parts of the spectrum is similar, only the avoided
crossings for the odd energy levels are narrower. The odd parity energy levels
correspond to wave functions vanishing
in the center of the tunnel barrier so the tunnel coupling between the rings is naturally
less pronounced than for the even energy levels.

\begin{figure*}[ht!]
\centerline{\hbox{\epsfysize=50mm
               \epsfbox {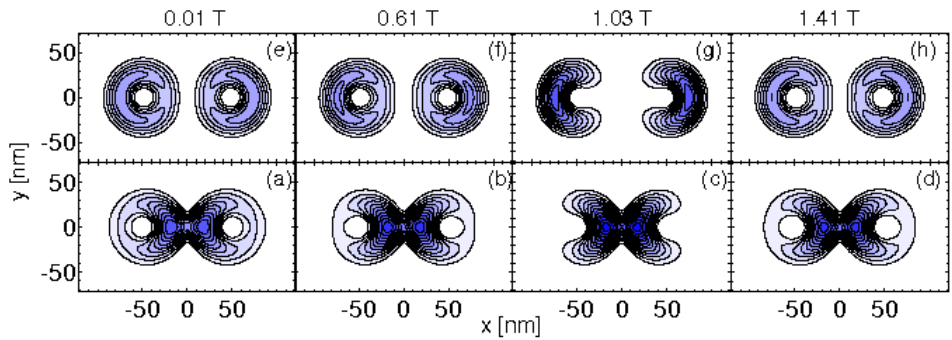}\hfill}}
\caption{Lower row of plots (a-d) shows the charge density in the
ground-state (of the even parity) for $d=3.25$  (the energy levels
are given in Fig. \ref{1eip}(c)). The upper row of plots shows the
first excited state, which is the lowest-energy state of the odd
parity. Columns correspond to different magnetic fields.
 \label{1eff}}
\end{figure*}

In the absence of the tunnel coupling between the rings  ($d\ge
3.75$) the electron density of the stationary states reproduces the
circular symmetry of separate rings --  irrespective of the
magnetic field value. A distinct dependence on the magnetic field
occurs only when the tunnel coupling is present.
 The charge
densities for the lowest even and odd energy levels are given for
$d=3.25$ in the lower (a-d) and upper (e-h) rows of Fig. \ref{1eff},
respectively. The lowest even (odd) energy level is the ground state
(the first excited state). In the ground state the electron tends to
stay near the symmetry center of the double ring system. The extent
of its localization varies with the magnetic field. The strongest
localization near the interring contact area [cf. Fig. \ref{1eff}(c)]
corresponds to the center of the avoided crossing of the two lowest
energy levels [see the spectrum of Fig. \ref{1eip}(c)]. In the odd
parity energy level the electron is by the symmetry forbidden to be
found at the center of the structure and it
tends to occupy the extreme ends of rings. The strength of
localization at the ends oscillates with the magnetic field and is the largest
near the center of the avoided crossings occurring between two lowest odd energy levels [see Fig. \ref{1eff}(g) for
$B=1$ T and Fig. \ref{1eip}(c)].

\begin{figure}[ht!]
\centerline{\hbox{\epsfysize=42mm
               \epsfbox {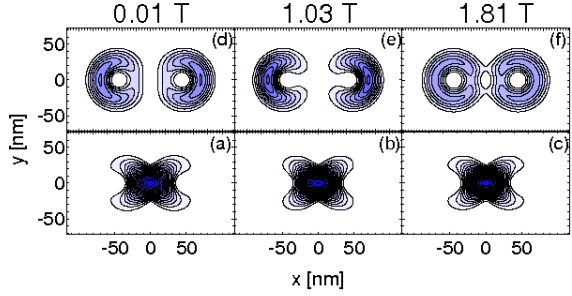}\hfill}}
\caption{Same as Fig. \ref{1eff} but for $d=3$. The corresponding
spectrum is displayed in Fig. \ref{1eff}(d).}
 \label{1effd3}
\end{figure}

Coalesced rings at $d=3$ form a cavity of increased width near $x=0$
(see Fig. 1),  which
acts like a quantum dot -- in the ground-state the electron
becomes localized  at the interring contact area [Fig. \ref{1effd3}(a-c)] and only weakly penetrates the more
distant parts of the structure. The ground-state electron density
weakly depends on the magnetic field since the electron is localized
within an area which is quite small.
The energy of this (ground) state is distinctly lowered with respect
to the excited part of the spectrum [compare Fig. \ref{1eip}(d) for $d=3$
Fig. \ref{1eip}(c) and $d=3.25$]. The excited energy levels are essentially
unchanged with respect to $d=3.25$, only the avoided crossings
become larger. Fig. \ref{1effd3} shows that the densities in the
first excited state of the coalesced system are similar to the
$d=3.25$ case: they stay spread all over the double ring structure
and are not limited to the contact area as in the ground-state.
 In the spectrum calculated for $d=3.25$ [Fig. \ref{1eip}(c)] a trace of
the Aharonov-Bohm oscillation for the ground-state energy is still
visible since at least outside the center of the avoided crossings
the ground state wave function encircles the rings (see Fig. \ref{1eff}).
For $d=3$ even this residual oscillation of the ground-state energy
disappears [see Fig. \ref{1eip}(d)]. However, oscillations are
still present in the excited part of the spectrum in which the wave
function covers the entire structure. Therefore the coalescing of
the rings mainly perturbs the ground state which becomes localized
in the quantum dot formed at the contact of the rings and not the
excited states which are spread all over the double ring structure.

\begin{figure*}[ht!]
\centerline{\hbox{\epsfysize=50mm
               \epsfbox {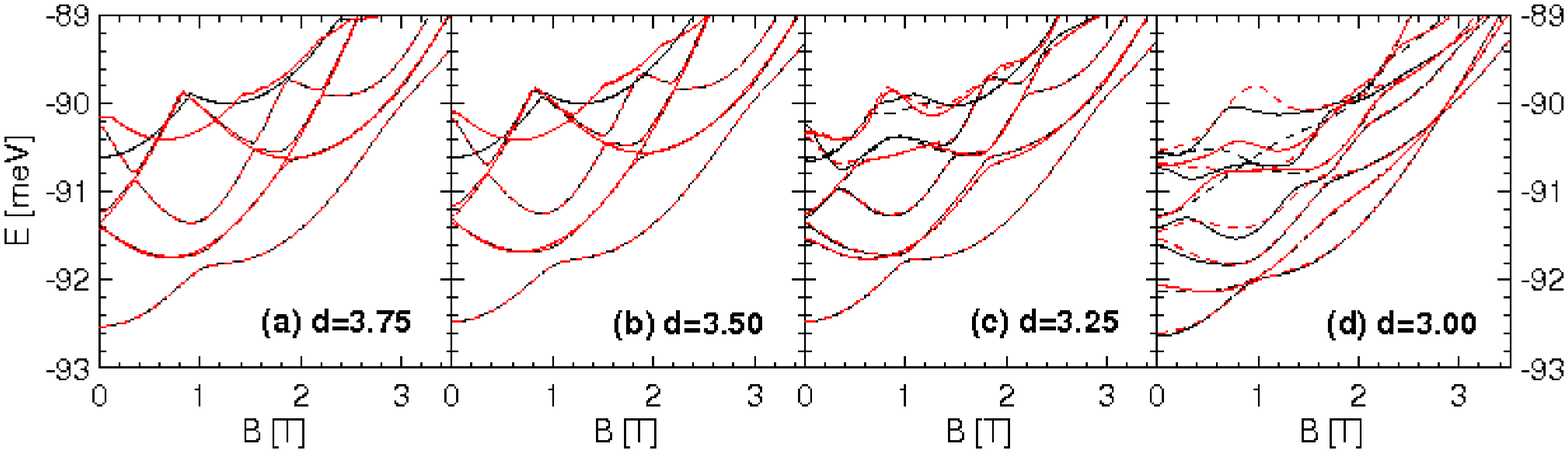}\hfill}}
\caption{Two-electron spectra for a pair of identical rings. Solid
and dashed curves show the even and odd parity energy levels. Black
and red lines correspond to the total spin $S=0$ and $S=1$.
 \label{2eip}}
\end{figure*}

\begin{figure}[ht!]
\centerline{\hbox{\epsfysize=70mm
               \epsfbox {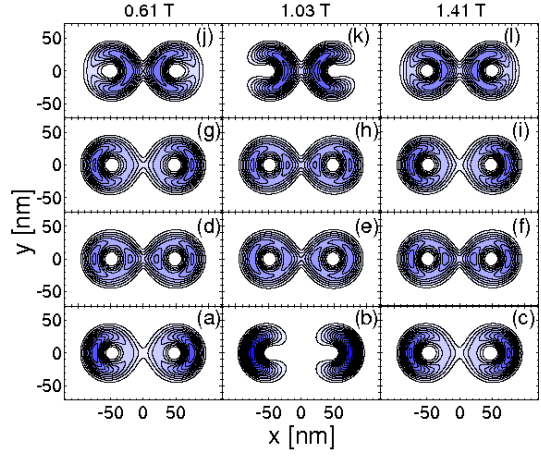}\hfill}}
\caption{Two-electron densities for $d=3.25$ for four lowest-energy
singlet states (for triplets the results are nearly identical). The
lower the plot the lower the energy. The ground state (a-c) and the
third excited state are of the even parity for all $B$. For $B=1.03$
T the first excited state is of the even parity and the second
excited state is odd. For the two other magnetic field the
symmetries of this two energy levels are inverted -- see also Fig.
\ref{2eip}(c). }\label{2ewff}
\end{figure}

\begin{figure}[ht!]
\centerline{\hbox{\epsfysize=70mm
               \epsfbox {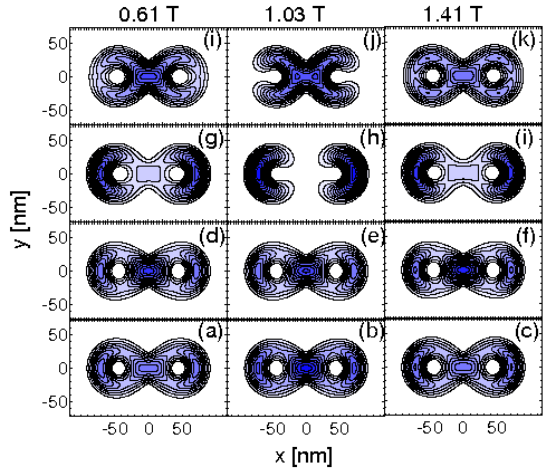}\hfill}}
\caption{Two-electron densities for $d=3$ for four lowest-energy
singlet states. The lower the plot the lower the energy. Plots (a-c)
correspond to the ground-state.} \label{mxp}
\end{figure}

\subsection{Identical rings: two electrons}
The electrostatic coupling between the rings appears with the second electron introduced
to the system.
The two--electron spectrum for a pair of rings is displayed in Fig.
\ref{2eip}. For two electrons the Coulomb repulsion makes the interring
barrier higher reducing the tunnel coupling and in the low--energy part spectrum the electrons tend to
occupy separate rings. For wide interring barrier and vanishing
tunnel coupling the electron separation is complete. In consequence ,
for $d=3.75$,
3.5 and 3.25 the ground state is two-fold degenerate [see Fig.
\ref{2eip}(a-c)]. One of the ground states is the spin singlet of
the even parity and the other is the spin triplet of the odd parity.
In the two-electron systems the singlet-triplet degeneracy (vanishing exchange interaction \cite{ei}) occurs when the
electrons occupy distinct and separated locations. For double dots this degeneracy is found for wide \cite{ei} interring barriers or high magnetic fields.
Similar is the effect of the Wigner crystallization in elongated, quasi one-dimensional quantum dots.\cite{1dqd}

For a single-electron the effects of
the tunnel coupling -- splitting of the odd and even parity energy
levels and avoided crossings in the odd and even parts of the
spectrum --  were visible already for $d=3.5$ [Fig. \ref{1eip}(b)]
and for $d=3.25$ [Fig. \ref{1eip}(c)] they were already quite
strong. Lifting of the spin
degeneracy is only visible for the coalesced rings [$d=3$ see Fig
\ref{2eip}(d)]. Note that also in this case the singlet and triplet
becomes degenerate when the magnetic field is switched on enhancing
the electron localization in opposite ends of the structure.

For wide interring barrier, when the tunnel coupling is absent, the
separate rings are only coupled by the Coulomb interaction of
confined electrons. The electrostatic potential of the electron
localized in one of the rings perturbs the symmetry of the other
ring. In consequence the angular momentum of a single ring is not a
good quantum number and one obtains avoided crossing between energy
levels [Fig. \ref{2eip}(a-c)] instead of the crossings [compare Fig.
\ref{1eip}(a)]. Note, that the center of the avoided crossing for
two electrons [see Fig. 2 \ref{2eip}(a)] occurs for the value of the
magnetic field for which the ground state angular momentum
transition occurs for a single electron [cf. Fig. \ref{1eip}(a)].

Electron densities are displayed in Fig. \ref{2ewff} for $d=3.25$.
In the ground-state the electrons stay at the opposite ends of the
rings. The most pronounced separation occurs at the center of the
avoided crossing near $B=1.03$ T. Oscillation as pronounced as in
the ground state is observed for the third excited singlet state.
In this state the electrons are mostly
localized in the center of the structure in contrary to the
ground-state, for which electron mostly occupy ends of the double structure. The
electron density in the third excited state is in fact similar to the
single-electron ground state [cf. Fig. \ref{2ewff}(j-l) and Fig.
\ref{1eff}(b-d)]. However, the Coulomb repulsion makes the
two-electron density less strongly localized.
In the first two excited singlets [Fig. \ref{2ewff}(d-i)] the reaction of the density to the
field is much weaker than for the ground-state and the third excited singlet.

The singlet electron densities for coalesced rings ($d=3$) are
displayed in Fig. \ref{mxp}. Comparison of this plot with the
single-electron densities of Fig. \ref{1effd3} indicates that in the
two lowest-energy two-electron levels [Fig. \ref{mxp}(a-f)]
the single-electron ground-state [with density localized
in the interring contact area -- Fig. \ref{mxp}(a-c)] and the first
excited state [electron density localized at the extreme ends of both the rings -- Fig. \ref{mxp}(d-f)] contribute nearly equally.
In the second [third]
excited state the single-electron first excited [ground] state
contribution is dominant -- see Fig. \ref{mxp}(g-i) [Fig. \ref{mxp}(j-k)]. Note that the type of the electron
localization observed here in the second-excited state [Fig.
\ref{mxp} (g-i)] corresponds to the two-electron ground state for the larger
value of $d=3.25$.

\begin{figure*}[ht!]
\centerline{\hbox{\epsfysize=80mm
               \epsfbox{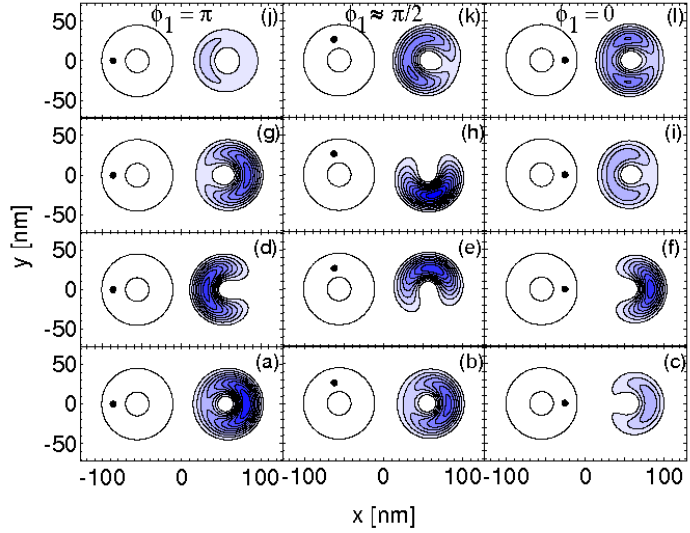}\hfill}}
\caption{Pair correlation function plots for spin singlets with the position of the electron in the left ring
fixed in the position marked by the dot. The distance parameter is $d=3.75$ and the magnetic field $B=0.51$ T. Lower rows of plots correspond to lower energies.} \label{pcf}
\end{figure*}

Although in all the low-energy states discussed here for $d\geq 3.5$
the electrons occupy different rings, in fact the type of the
interring correlation between the electron positions varies
significantly from state to state. Fig. \ref{pcf} shows the pair
correlation function for spin singlet states calculated at $d=3.75$
and $B=0.51$ T for the ground state [Fig. \ref{pcf}(a-c)], and the excited states
[Fig. \ref{pcf}(d-l)]. In each of the three columns we assumed that the electron
in the right ring is localized in a different position -- marked by the dot in Fig. \ref{pcf}.
In the ground-state the electron in the right ring tends to
stay far from the left ring and the angular reaction of the electron
in the right ring on the actual position of the the electron in the
left ring is weak [Fig. \ref{pcf}(a-c)]. Only a slight rotation of the electron probability distribution
in the left ring is
observed in Fig. \ref{pcf}(b) for the electron in the left ring localized at $\phi=\pi/2$ angle. For the ground state the overall PCF
values within the right ring are reduced when the electron in the
left ring is localized closer to the right ring [see Fig. \ref{pcf}(a-c) -- contour scale
is kept the same in all the plots]. In the third
excited state an opposite tendency is observed [see Fig. \ref{pcf}
(j-l)]: the overall value of the electron probability distribution {\it increases} when the electron in the
left ring approaches the right ring. In this state the minimal PCF value
within the right ring is found always near $\phi=0$. In contrast to the ground state and the third excited
state a strong correlation in the angular positions of the electrons
is observed for the first [Fig. \ref{pcf}(d-f)] and second [Fig.
\ref{pcf}(g-i)] excited states.

In the first excited state the
electron in the right ring stays away of the left ring when the
electron in the left ring is at $\phi=0$ [Fig. \ref{pcf}(f)], which
is natural for repulsing particles. However, when the electron in
the left ring is localized above its center $\phi=\pi/2$, the
electron in the right ring is also most probable to be found above
the center of its ring  [Fig. \ref{pcf}(f)] near
$\phi=\pi/2$, although one should rather expect a maximum of the probability density
at the opposite side of the center. Also in the second excited state the correlation is
somewhat different from what one might expect of the repulsing electrons.
For instance when the electron in the left ring is localized at the
closest distance to right ring the electron in the left ring tends
to approach the left ring [Fig. \ref{pcf}(g-i)].

In order to explain the observed features of the electron-electron
correlation in the four-lowest energy singlet states one needs to
consider the two-electron wave function. Below we present an approximate
analysis for $B\simeq 0.5$ T, when the total
wave function is mainly constructed of the single-ring
single-electron states of $l=0$ and $l=1$ angular momenta [see Fig.
\ref{1e1p}], i.e. the single-electron ground state and the first
excited state. Other energy levels lie much higher in the energy. The weak reaction of the electron in the right ring
to the angular position of the electron in the left ring observed in
the ground-state indicates that the two-electron wave function is
nearly separable. The spatial part of the singlet ground-state wave
function can be written in a following approximate non-normalized
form
\begin{eqnarray}
\Psi&=&(1+ P_{12})\chi_l(r_1)\left[f_0^l(\phi_1)-c f_1^l(\phi_1)\right]\nonumber \\
&&\times \chi_r(r_2)\left[f_0^r(\phi_2)+c f_1^r(\phi_2)\right],
\label{twv}
\end{eqnarray}
where $P_{12}$ is the electron exchange operator,\cite{opis}
$\chi_a$ is the radial wave function of the confinement within the
ring $a$ ($a=l$ for the left and $r$ for the right ring), and
$f_{m}^a$ is the angular momentum $m$ eigenstate for the electron
localized in the ring $a$ $f_m(\phi)=\exp(im\phi)/\sqrt{2\pi}$. The
argument of $f_m^a$ is the angular coordinate measured with respect
to the center of ring $a$. In Eq. (\ref{twv}) the real valued $c\geq
0$ introduces mixture of $l=1$ to $l=0$ eigenstates. Mixture of
these states is no longer rotationally invariant. For wave function
(\ref{twv}) the electron in the left ring is localized
preferentially near $\phi=\pi$ and the electron in the right ring
near $\phi=0$. For spatially separated $\chi_l$ and $\chi_r$ radial
functions, the two-electron density is given by
\begin{eqnarray}
|\Psi|^2&=&(1+P_{12})|\chi_l(r_1)\chi_r(r_2)|^2\left|f_0^l(\phi_1)-c
f_1^l(\phi_1)\right|^2\nonumber
\\ &&\times
\left|f_0^r(\phi_2)+cf_1^r(\phi_2)\right|^2.
\end{eqnarray}
For the first electron fixed within the left ring, $\chi_r(r_1)\rightarrow 0$ and
$P_{12}$ may be skipped of the above formula since the permutated
term vanishes.\cite{opis} We are
left with the density which is a separable product of single-electron densities 
whose angular dependence on $\phi_2$ becomes independent of
$\phi_1$, although the overall pair correlation function value
decreases when $\phi_1$ approaches 0 (the right ring), in agreement
with Fig. \ref{pcf}(a-c).

Now let us turn our attention to the third excited state. In this
state the electrons tend to occupy the area of the contact between
the rings and the PCF shows a weak angular correlation between the
electrons [Fig. \ref{pcf}(j-l)] like in the ground state. The wave
function which produces these properties differs from Eq.
(\ref{twv}) by the sign of $c$
\begin{eqnarray}
|\Psi|^2&=&(1+P_{12})|\chi_l(r_1)\chi_r(r_2)|^2\left|f_0^l(\phi_1)+c
f_1^l(\phi_1)\right|^2\nonumber
\\ &&\times
\left|f_0^r(\phi_2)-cf_1^r(\phi_2)\right|^2.
\end{eqnarray}
for which the electron localization angles are inverted (shifted by $\pi$) with respect to the ground state.

In the ground state and in the third excited state the $l=0$ and
$l=1$ single-electron angular eigenstates are mixed {\it within each of the rings} and the resulting probability densities are (nearly) separable. In the
first and second excited states the  angular momenta $l=0$ and $l=1$ eigenstates
contribute in a different manner. In these states the angular
momentum of the electron in one of the rings is $l=0$ while the
angular momentum of the electron in other ring is $l=1$. This case is described by the following wave function
\begin{eqnarray}
\Psi&=&(1+P_{12})\left[\chi_l(r_1)f_0^l(\phi_1)\chi_r(r_2)f_1^r(\phi_2)\pm\chi_l(r_1)f^l_1(\phi_1)\chi_r(r_2)f_0^r(\phi_2)
\right] \label{pipi}
\end{eqnarray}
(for brevity we concentrate on the spin-singlets only). At $B=0.51$
T for $d=3.75$ the first (second) excited singlet is of the odd
(even) parity, which corresponds to the '$+$' ('$-$') sign in the
above formula. In the absence of the overlap between the single-ring wave functions
the two-electron density for the first electron fixed in the left ring is given
by (up to the normalization constant)
\begin{eqnarray}
|\Psi|^2\rightarrow |\chi_l({ r}_1)\chi_r({ r}_2)|^2\left(1\pm
\cos(\phi_1-\phi_2)\right), \label{form}
\end{eqnarray}
where $\phi_1$ is measured with respect to the center of the left
ring and $\phi_2$ with respect to the center of the right ring.
Formula (\ref{form}) indicates that in the first excited state (of
the odd parity) both the electrons are most probably localized at
the same angle $\phi_1=\phi_2$, and in the second excited state (of
the even parity) the electrons are localized at opposite angles
$\phi_1=\phi_2+\pi$, which explains the behavior observed in Fig.
\ref{pcf}(g-i). At a close inspection in Fig. \ref{pcf}(h) one notices
that the exact angular position $\phi_2$ of the maximum in the right
ring deviates off $\phi_1$ a little bit, which is due to the
contribution of the higher angular momenta to the wave functions. We see that in contrast
to the ground-state and the third excited state the angular
correlations between the electrons are strong in the first and
second excited states. According to the presented arguments the
symmetrization of the wave function by $(1+P_{12})$ operator [as well as the antisymmetrization by $(1-P_{12})$] does
not influence the correlated properties of the system when the
electrons are spatially separated. Then, {\it i}) the singlet and triplet energy levels become degenerate and {\it ii}) the type of the correlation depends
only on the form of the original wave function prior to symmetrization by $(1\pm P_{12})$. The wave
function of the first and second excited states that is symmetrized
in [Eq. (\ref{pipi})] is essentially entangled, hence the strong
interring angular correlation. For the ground-state and the third excited
state with negligible angular correlations the symmetrization was
performed on a strictly separable product of single-electron wave
functions.

\begin{figure*}[ht!]
\centerline{\hbox{\epsfysize=50mm
               \epsfbox{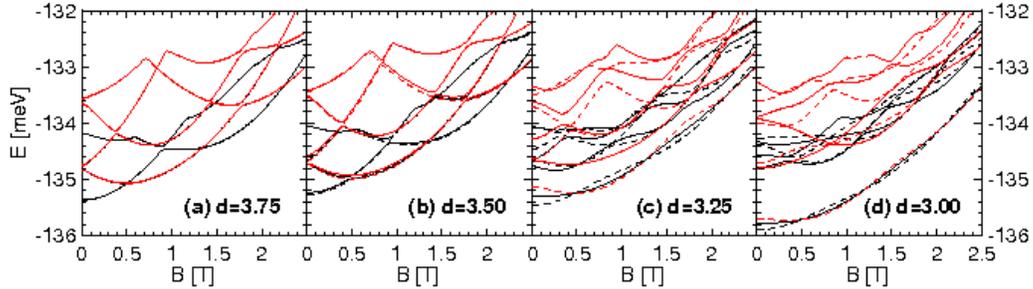}\hfill}}
\caption{Three-electron spectra for a pair of identical rings. Solid
and dashed curves show the even and odd parity energy levels. Black
and red lines correspond to the total spin $S=1/2$ and $S=3/2$.
\label{3eip}}
\end{figure*}

\subsection{Three electrons in a pair of identical rings}
The  spectrum for three electrons at wide interring barrier is given
in Fig. \ref{3eip}(a). In the ground-state energy level at $d=3.75$
we notice crossings of energy levels related to symmetry transformations which were
also present in the single-electron spectrum but which were absent for two-electrons.
We notice that the "period" of the
ground-state energy level oscillation is halved with respect to the
single ring case [cf. Fig. \ref{1eip}]. In the low-energy spectrum
we have two electrons in one ring and a single electron in the
other. Halving of the oscillation period results of the fractional
Aharonov-Bohm effect \cite{czakra} occurring for few-electron states
confined in a single ring. The ground-state oscillations are due to
the spin-transformations of the electron pair in a single ring. The angular momentum of
the two-electron subsystem is not a good quantum number due to the perturbation of the two-electron ring
by the potential of the electron in the right ring.
For
$B<0.5$ T the ground state is two-fold degenerate: with respect to
the parity and corresponds to $S=1/2$. For low magnetic fields the
spins of the electrons in the single ring are opposite and
compensate, so the total spin can be identified with the spin of the
solitary electron in the other ring. For $B>0.5$ T the ground state
of the two-electron subsystem is the spin-triplet. Since the spin of the
solitary electron may have an arbitrary orientation the ground-state
becomes four-fold degenerate: with respect to both parity and the
both allowed total spin quantum number values $S=1/2$ and $3/2$. For
smaller $d$ the degeneracies are lifted [see Fig. \ref{3eip}(b-d)].
Near the ground-state we notice a characteristic oscillation of the
ground-state symmetry [Fig. \ref{3eip}(c,d)]. The ground-state is
mostly either the odd-parity low-spin $S=1/2$ state or the
odd-parity spin-polarized state $S=3/2$. When the energy order of
these two energy levels changes they become nearly degenerate with a
third state: the even-parity low-spin state. The even-parity
spin-polarized ($S=3/2$) state runs much higher in the energy. This
sequence of the ground-state spin-- and parity--symmetry
oscillations is also characteristic to three-electron
circular\cite{mikhailov,3e,li} dots [for circular dots states of angular momentum quantum
number which is even (odd) integer are of the even (odd) parity] as well as to elliptic\cite{li} quantum dots and double quantum dots \cite{3e} containing three electrons.
For elliptic quantum dots the near degeneracy of the three energy
levels that we obtain here for discrete values of the magnetic field indicates
that the  deformation of the confinement potential with respect to
the circular limit is strong (see the discussion given in Ref. \cite{li} for elliptic dots). For less strong deformation
the even parity low-spin state becomes a ground-state for some
narrow but distinct magnetic field ranges. \cite{3e,li}

\subsection{Oscillations of the three-electron charge density with the magnetic
field}

\begin{figure}[ht!]
\centerline{\hbox{\epsfysize=50mm
               \epsfbox{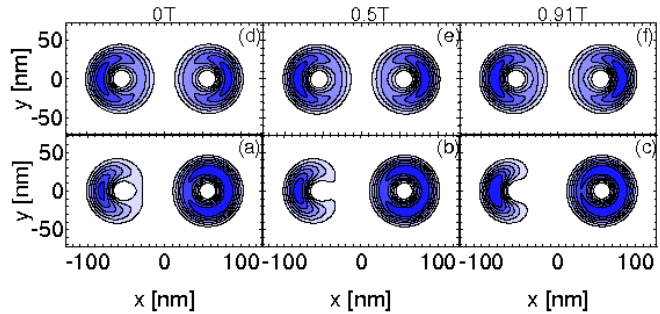}\hfill}}
\caption{Ground-state charge density for three electrons in
identical rings (d-f) and for the right ring deeper by 0.1 meV (a-c)
at the interring spacing of $d=3.75$.} \label{nier}
\end{figure}

The results for the charge density of one- and two- electron systems
presented above exhibited a distinct magnetic field dependence. The
results for three electrons at $d=3.75$ presented in Fig.
\ref{nier}(d-f) indicate that the dependence on the magnetic field
is significantly weaker.
 For ideally symmetric pair of rings the charge
density of both even and odd parity three-electron eigenstates is
distributed equally between the rings with 1.5 electron charge per
ring on average.
 A classical
distribution with one electron in one ring and two in the other can only
occur when the symmetry is lifted. In the absence of the tunnel
coupling a potential well difference of 0.1 meV is enough to obtain the integer
distribution of electrons between the rings. The charge density for
$d=3.75$ and the right ring deeper by 0.1 meV is shown in Fig.
\ref{nier}(a-c). The dependence of the single-electron charge
localized in the left ring on the magnetic field is clear, and the
reaction of the two-electron density in the right ring is weaker.
Note that in this plot the deviation of the single-electron density
off the circular symmetry occurs for $B=0.91$ T, for which the
two-electron density is closest to circular, which leads to a
compensation of the magnetic oscillation of the density for
identical rings.

\begin{figure}[ht!]
\centerline{\hbox{\epsfysize=40mm
               \epsfbox{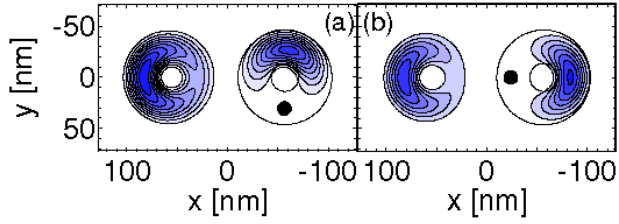}\hfill}}
\caption{Pair correlation function for the ground-state at $B=0$ and
$d=3.75$ for the right ring deeper by 0.1 meV than the left one.
Left ring contains a single electron and the right one two
electrons. A position of one of the electrons in the right ring is
fixed and marked by the dot. Contours of the ring area are also
shown.} \label{pcf3e}
\end{figure}

\begin{figure}[ht!]
\centerline{\hbox{\epsfysize=60mm
               \epsfbox {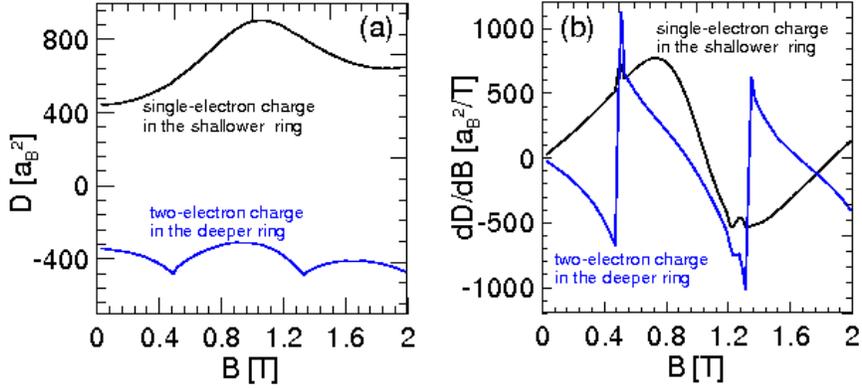}\hfill}}
\caption{(a) Deviations o the single-ring electron densities off the
circular symmetry [Eq. (\ref{o})] in function of the magnetic field
for the system of three electrons in double ring structure with the
right ring deeper by 0.1 meV for $d=3.75$. Blue curve corresponds to
the deeper ring occupied by two electrons and the black curve to the
shallower ring which contain a single electron. (b) Derivative of
$D$ over $B$. $a_B$ is the Bohr radius.} \label{odchy}
\end{figure}

In order to quantify deviations of the single-ring electron density
off the circular symmetry we calculated the parameter
\begin{equation} D=<(x-X)^2>-<(y-Y)^2>,\label{o}\end{equation} where $X$ and $Y$ are coordinates
of the ring and the average is calculated over the charge density of
separate rings. The result is displayed in Fig. \ref{odchy} with the
blue curve for the two-electron density of the deeper ring and with
the black curve for the single-electron density of the shallower
ring. The single-electron parameter is larger, varies more strongly
with the magnetic field and its variation is continuous in contrast
to the two-electron value, which has cusps when the ground-state
spin state of the two-electron subsystem changes. Moreover, the
single-electron parameter is positive which indicates that the
deformation occurs rather in the horizontal ($x$) direction, while
the two-electron deviation occurs mainly in the vertical direction
($y$). The absolute value of the deviation of the two-electron
density from circular is the largest at the symmetry
transformations. As noted in context of Fig. \ref{nier} the largest
deformation of the single-electron density corresponds to a weak
deformation of the two-electron density.

When the two-electron subsystem changes its spin state the ground
state charge density in the deeper ring is modified in a
discontinuous manner. Due to the Coulomb coupling this change may
influence the electron density of the shallower ring, which might react in a discontinuous manner.
 In order to
quantify this reaction we calculated derivatives of the $D$
parameter with respect to the magnetic field. The derivative of the
two electron parameter is discontinuous at the cusps [see Fig.
\ref{odchy}(b)]. We notice that the single-electron parameter
exhibits an irregular structure when the two-electron density
changes. However, this structure is not very pronounced. This result
along with the presented above reference calculations indicate that
the charge density of each ring is quite indifferent to the actual
form of the charge distribution in the other ring. Additional argument for that
conclusion is provided by the pair correlation function.

The pair correlation function plot for three electron system with two
electrons in the right ring is shown in Fig. \ref{pcf3e} for the
ground-state at zero magnetic field. We fix a position of one of the
electrons in the right ring and mark it with the dot in the figure.
We notice that the angular distribution of the solitary electron in
the left ring is unaffected by the position of the fixed electron.
 Only the electron in the
same (right) ring reacts to the variation of the fixed electron
position [compare Fig. \ref{pcf3e}(a) and \ref{pcf3e}(b)].  Note,
that for the electron fixed below the center of the right ring [Fig.
\ref{odchy}(a)] the other electron in the right ring is not exactly
on the other side of the center. It tends to avoid the left part of
the ring and the position of the maximum is localized below the
angle $\phi=\pi/2$. The result of Fig. \ref{pcf3e} indicates that in
rings separated by a barrier that is thick enough to
prevent the interring tunneling  the electrons are mainly coupled as
entire charge densities and do not react on their actual position. A
strong angular correlation is only observed for the electron within
the same ring.

\subsection{Magnetization}

\begin{figure*}[ht!]
\centerline{\hbox{\epsfysize=50mm
               \epsfbox {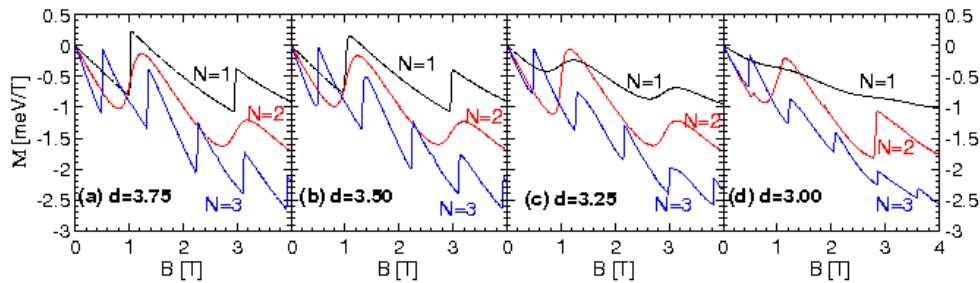}\hfill}}
\caption{Magnetization of a pair of identical rings filled with 1, 2
or 3 electrons plotted with black, red and blue curves,
respectively.} \label{ma}
\end{figure*}

Fig. \ref{ma} shows the magnetization $M=-\frac{\partial E}{\partial
B}$ of the pair of identical rings filled by 1, 2 and 3 electrons
with different spacing parameters $d$. For a single circular ring
the magnetization is discontinuous at the angular momentum
transitions. The interring tunnel coupling makes the single-electron
magnetization smooth and continuous [see black curves in Fig.
\ref{ma}(a-d)]. The oscillation of $M$ with $B$ are extinguished
for the coalesced rings, when the Aharonov-Bohm oscillations occurs
only in excited states and not in the ground-state which is
localized in the quantum dot formed at the contact of the rings.

 For two electrons the magnetization
is a smooth function of the magnetic field unless the rings are
close enough to form a single-structure. Two-electron ground-state
symmetry transformations which produce the cusps in the
magnetization occur only for the coalesced rings ($d=3$). For three
electrons the symmetry transformations are present for any $d$. For large
interring distance they correspond to the spin transformations of
the subsystem of two electrons confined within the same ring. The
spin transformations of the three electron system occur also when the rings form a single
coalesced structure.

\subsection{Charging properties}

For a structure embedded in a charge tunable device the charging of
the double ring by $N-th$ electron occurs when the chemical
potential $\mu_N=E_N-E_{N-1}$ (where $E_N$ is the ground-state
energy of $N$ confined electrons) is aligned with the Fermi level of
the electron reservoir. In capacitance spectroscopy \cite{lorke} the cusps of the charging lines
in function of the magnetic field indicate symmetry
transformations of the confined system.

\begin{figure*}[ht!]
\centerline{\hbox{\epsfysize=50mm
               \epsfbox{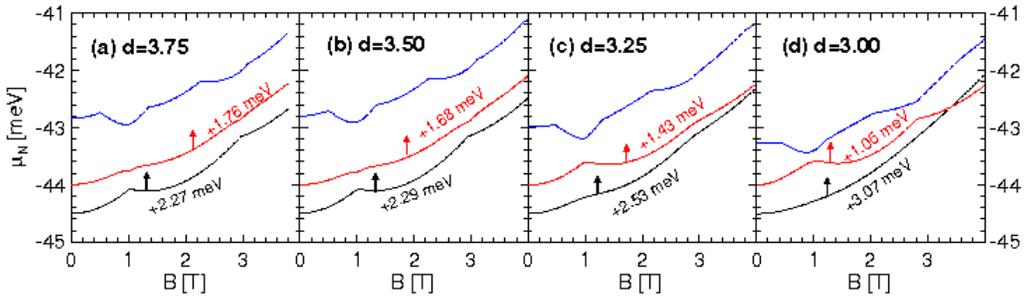}\hfill}}
\caption{Chemical potential of 1, 2 and 3 electrons confined in the
double ring plotted with black, red and blue curves, respectively.} \label{cz}
\end{figure*}

Fig. \ref{cz} shows the calculated chemical potentials in function
of the magnetic field. Chemical potentials of 1 and 2 electrons were
shifted up on the energy scale for clarity (the amount of the shift
is given in the figure). In the absence of the tunnel coupling [Fig.
\ref{cz}(a,b)] the chemical potential of two electrons is a smooth
function of the magnetic field with exception to the V shaped cusps
that result of the single-electron angular momentum transitions that
produce the $\Lambda$ shaped cusps on the single-electron chemical
potential. Since the two-electron ground-state energy is smooth as a
function of $B$ all the cusps of the three electron system have the $\Lambda$
shape and result of the symmetry transformations of the three
electron system. The parabolic minimum of $\mu_3$ between the
$\Lambda$ cusps is due to the smooth maximum of $E_2$ that occurs at
the avoided crossings near odd multiples of flux quanta (see Fig.
\ref{2eip}). Due to the fractional Aharonov-Bohm oscillation of the
two electron subsystem we find two $\Lambda$ cusps in $\mu_3$ for a
single one in $\mu_1$.

For $d=3.25$ the $\Lambda$ cusps of the single-electron chemical
potential are smoothed out and in consequence the V cusps of $\mu_2$
disappear [Fig. \ref{cz}(c)]. The ground-state avoided crossing for
the two-electrons becomes narrower, so $\mu_2$ acquires a distinct
maximum near $1$ T. A $\Lambda$ cusp in $\mu_2$ and consequently a V
cusp in $\mu_3$ occur for the coalesced rings [Fig. \ref{cz}(d)].
The presented results indicate that the strength of the interring
coupling may be deduced of the charging experiments.

\subsection{Separability of the system at large interring barrier}

In the absence of the tunnel coupling the rings are coupled only
electrostatically. In order to answer the question to which extent
the distinct rings may be treated as separable we performed a reference
calculations in which we considered
a {\it single} ring perturbed by the Coulomb potential of
a classical point charge localized in the other ring.
 In the reference calculation for the two-electron system we
assumed that in the left (right) ring a classical point-charge
electron is localized at $\phi=\pi$ ($\phi=0$) and calculated the
single-electron spectrum for the confinement potential of the right
(left) ring. The two-electron spectrum of a ring couple was then
estimated by the sum of single-electron spectra of the left and right rings.

\begin{figure}[ht!]
\centerline{\hbox{\epsfysize=50mm
               \epsfbox {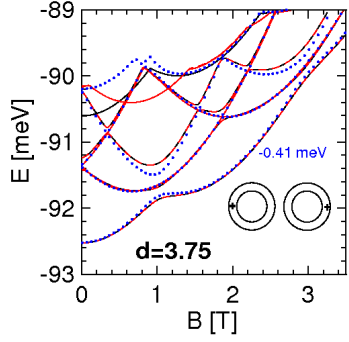}\hfill}}
\caption{The exact two-electron spectrum (lines) and the reference
spectrum (dots) obtained as a sum of single-electron spectra for a
single ring perturbed by a classical charge within the other ring.
The reference spectrum was shifted down by 0.41 meV on the energy
scale. The crosses in the inset show the assumed positions of the
classical charges.} \label{sepa}
\end{figure}

Comparison of the obtained result with the exact two-electron
spectrum is given in Fig. \ref{sepa}. The reference calculation was
shifted down by 0.41 meV to coincide with the exact two-electron
energy for $B=0$. We see that the magnetic-field dependence of the
ground-state energy and the nearly degenerate first excited energy level is quite
accurately described by the ansatz model. In fact the exact first
excited energy level is not exactly degenerate and there are two
energy levels for states of different parities. The reference
calculation in which the rings are treated separately overlooks this
splitting. Although the width of the exact avoided crossing near
$B=1$ T is quite accurately described by the reference calculation
the increase of the exact ground-state energy in the center of the
avoided crossing is smaller than in the reference calculation. The
higher part of the exact spectrum deviates off the reference
calculation significantly.

\begin{figure}[ht!]
\centerline{\hbox{\epsfysize=50mm
               \epsfbox {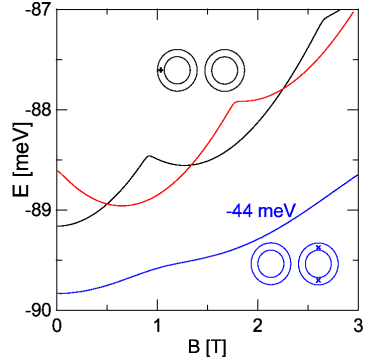}\hfill}}
\caption{Blue line shows the energy of the single electron (shifted
down by -44 meV) within the left ring with the Coulomb potential of
two classical point charges localized as indicated in the blue inset
to the figure. Black and red curves show the lowest-energy singlet
and triplet for two electrons in the right ring with a classical
point charge localized in the left ring in the position marked in
the black inset. Distance parameter between the rings is $d=3.75$
and the depth of the right ring is increased by 0.1 meV. }
\label{luzem}
\end{figure}

The reference spectrum for three electrons was calculated as a sum
of two spectra 1) the single-electron confined within the left ring
which is perturbed by two classical charges localized in the right
ring at positions marked by the crosses in the blue inset to Fig.
\ref{luzem} and 2) the two-electron spectrum of the right ring with
the classical point charge localized in the left ring in the
position of the cross in the black inset to Fig. \ref{luzem}. Blue
curve in Fig. \ref{luzem} shows the single-electron ground-state,
black and red curves correspond to the lowest-energy singlet and
triplet states of the two-electron ring, respectively.
For a single circular ring the confined electron system undergoes momentum transitions
as function of the magnetic field.  For the two-electrons the transitions are accompanied by the
spin transformations. When the two rings -- one containing a single electron and the other two electrons --
become electrostatically coupled the angular momentum transformations disappear.  However,
the spin ground-state transitions are still present in the
two-electron subsystem.

\begin{figure}[ht!]
\centerline{\hbox{\epsfysize=50mm
               \epsfbox{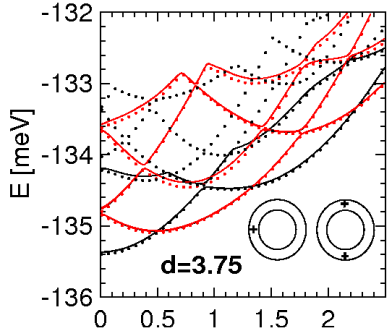}\hfill}}
\caption{The exact three-electron spectrum (lines) and the reference
calculation (dots). The reference spectrum is a sum of a
single-electron spectrum in the left ring perturbed by two classical
charges and the two-electron spectrum of the right ring with a
classical charge localized in the left ring (see the inset to this
figure and to Fig. \ref{luzem}). Black color corresponds to the
low-spin states and red color to the high spin states. The reference
spectrum was shifted down by 1.41 meV.} \label{sepa3}
\end{figure}

The reference spectrum for the three-electron system calculated using the
two separate calculations is compared to the exact spectrum in Fig.
\ref{sepa3}. In the ground-state both calculations agree quite
accurately, but differences are observed in the excited states.

\subsection{Electron-electron correlation in the coalesced double rings}
\begin{figure}[ht!]
\centerline{\hbox{\epsfysize=120mm
               \epsfbox{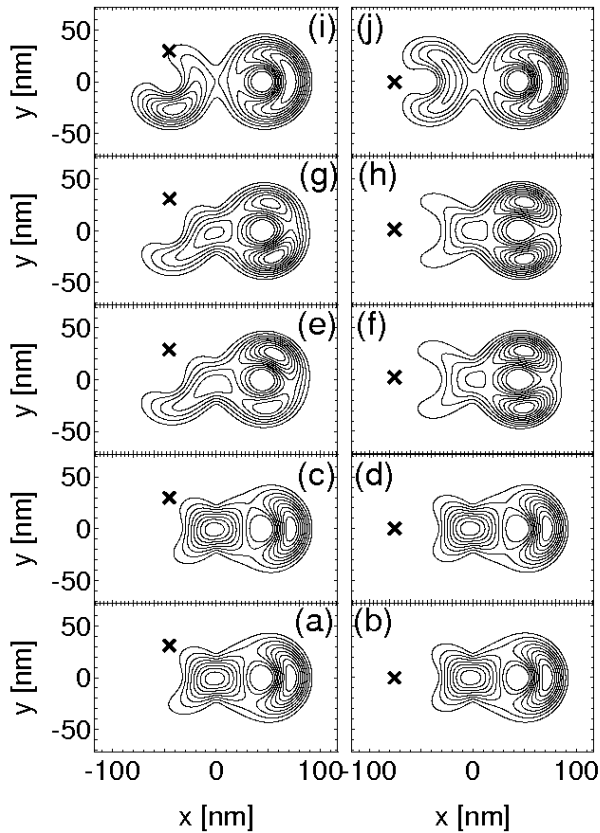}\hfill}}
\caption{Pair correlation function for three electrons in coalesced double ring structure $d=3$ at $B=0$.
Plots (a,b) correspond to the ground-state, and higher rows to subsequent {\it low spin} ($S=1/2$) energy levels.
The two columns correspond to two locations of the fixed position electron which is marked by the
cross in the figure.}\label{3pcfko}
\end{figure}

The pair correlation plots for the three electron system
presented so far were limited to large interring barrier. Let us look at the correlation
when the rings form a single structure.
The pair correlation function
plots for $d=3$ and $B=0$ are given in Fig. \ref{3pcfko} for the low-spin states.
Near the ground state one of the electrons occupy
the contact area between the rings and the two others are localized at the left
and right ends of the double structure. In the plots we present results for two
different locations of the electron in the left ring (marked by crosses in Fig \ref{3pcfko}).
The pair correlation for the lowest-energy odd parity state  Fig. \ref{3pcfko}(a,b) and
the first excited state Fig. \ref{3pcfko}(c,d) are similar.  Both these states have similar energies
(see the spectrum in Fig. \ref{3eip}(d)].
Only the electron localized at
the contact between the rings reacts to the position of the electron in the left ring, and this reaction
is not very pronounced [see Fig. \ref{3pcfko}(a-d)].
A stronger reaction is observed in the three next excited states [Fig. \ref{3pcfko}(e-j)],
which correspond to distinctly higher energies [Fig. \ref{3eip}(d)]. In the highest energy state
of the presented set the ''quantum dot'' formed at the contact is empty and the correlation properties
[Fig. \ref{3pcfko}(i,j)] are similar to the ground-state at large interring barrier [cf. Fig. \ref{pcf3e}].

Results presented above for wide interring barrier
indicated that in the ground-state the actual electron positions
are correlated only within the same ring. Fig. \ref{3pcfko} demonstrates
that for rings forming a single structure the electron-electron correlation in the ground state still has only
a short-range character.
In the studied case each of the three electrons occupy a different spatial location and is quite indifferent to the actual positions
of the other electrons within their charge islands.

\subsection{Rings of different radii}

\begin{figure}[ht!]
\centerline{\hbox{\epsfysize=50mm
               \epsfbox {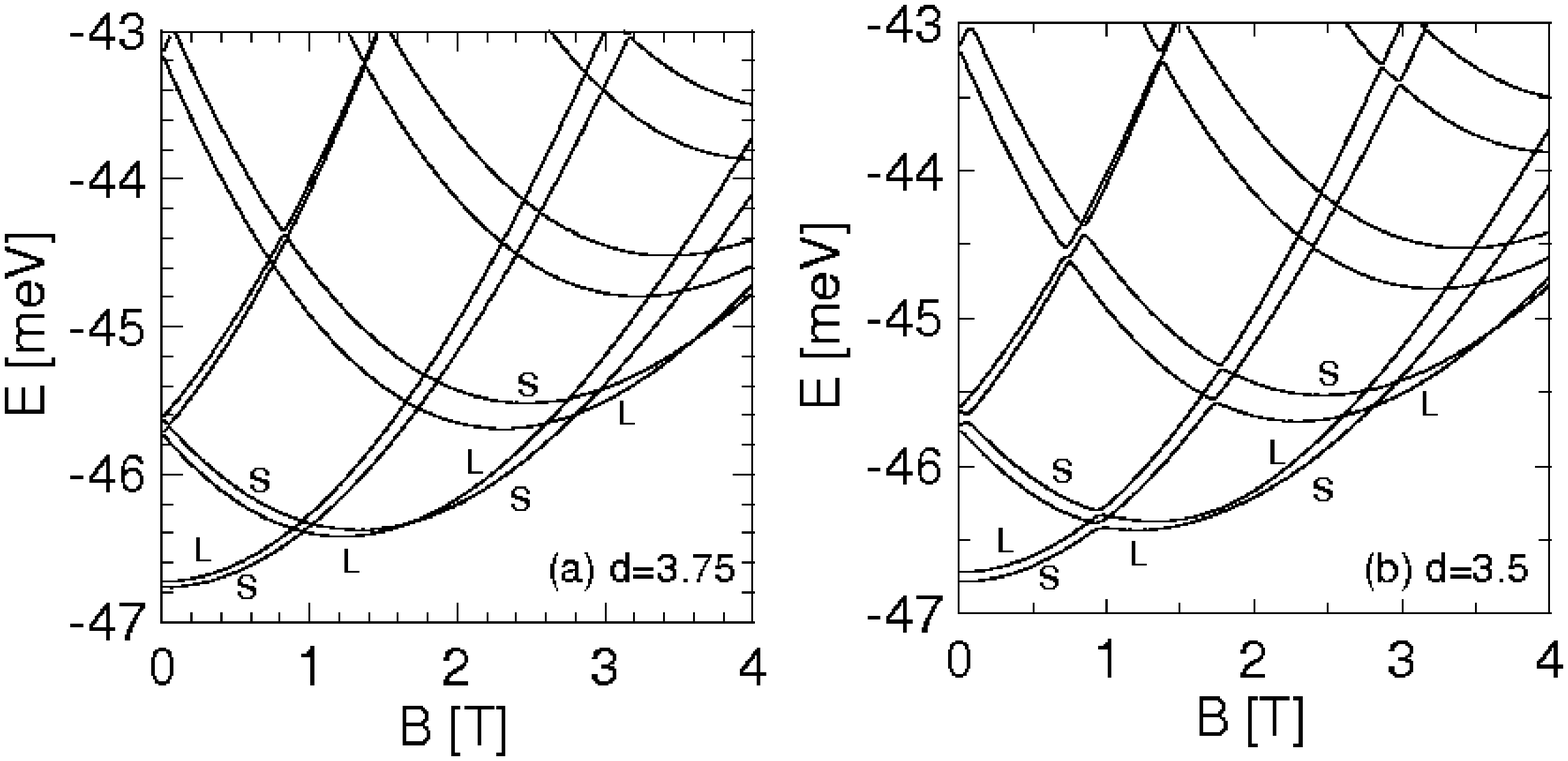}\hfill}\hbox{\epsfysize=50mm
               \epsfbox {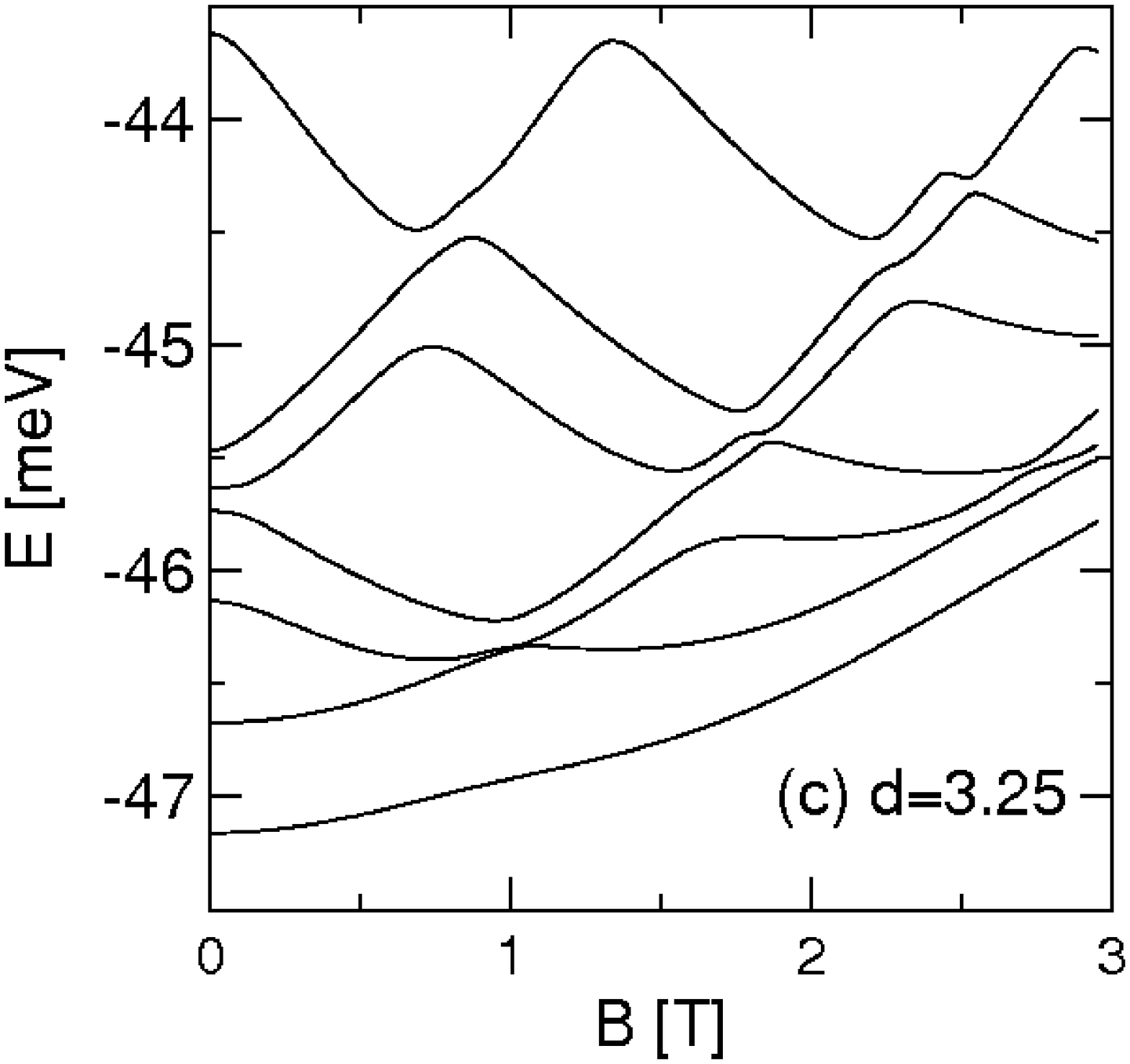}\hfill}}
\caption{Single-electron spectrum for a pair of rings of identical
depth but with the radius of the right ring increased by 5\%. In (a)
and (b) states localized mostly within the smaller (larger) ring are
labeled by S (L). In (c) electron in the ground state is nearly
equally distributed between the rings.} \label{widmo}
\end{figure}

Results presented so far were obtained under assumption that both
the rings have the same size. Let us now consider two rings that
possess slightly different radii. Namely, we assume that the radius of the right
ring is increased by 5 percent. The calculated single-electron
spectra are displayed in Fig. \ref{widmo}.

\begin{figure}[ht!]
\centerline{\hbox{\epsfysize=50mm
               \epsfbox {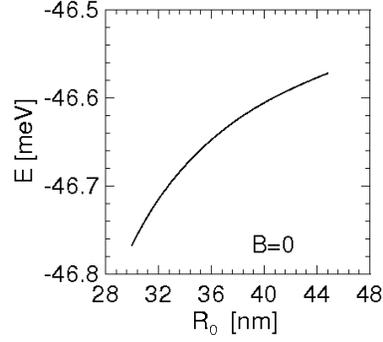}\hfill}}
\caption{Ground-state energy in function of the ring radius for a single electron within
a single ring.} \label{x}
\end{figure}

The energy of the lowest
zero angular momentum state in a single ring increases with
its increasing radius (see Fig. \ref{x}).
For that reason in the absence of the magnetic
field the zero-angular momentum ground-state is localized in the
smaller ring [energy levels marked by S in Fig. \ref{widmo}(a,b)] and the first excited state
is localized in the larger ring (energy levels marked by L in Fig. \ref{widmo}). On the contrary, due to
the centrifugal potential at zero magnetic field the states of the
non-zero angular momentum have lower energy when localized in the
larger ring [see Fig. \ref{widmo}(a)]. When the magnetic field is switched on the energy
levels that originate of the positive angular momenta single-ring
states go down in the energy due to the interaction of their
paramagnetic dipole moments with the external field [see the last term in the Hamiltonian (\ref{Eq1})] and they replace
the zero-angular momentum low-field ground-state when $B$ is high
enough. Consequently, for  $B$ exceeding 1 T the ground-state
becomes localized in the larger ring.  For $B>1.2$ T the ground
state (in the larger ring) and the first excited state (localized in
the smaller ring) start to increase in the energy. This increase
results of the diamagnetic shift [related to the
second term at right of (Eq. \ref{Eq1})]. It is a general rule that the
diamagnetic shift is stronger for more delocalized states (covering areas of larger  $\rho$).
 In
consequence one obtains another change of the electron localization: above $B=1.8$ T the ground state is localized again in the
smaller ring.  The ground-state transition to the larger ring occurs
again when the states originating of higher angular momenta become
ground-states. For $d=3.75$ the change in the ground-state
localization occurs through crossing of energy levels [see Fig.
\ref{widmo}(a)]. Already for $d=3.5$ tunnel-related avoided
crossings are obtained [see Fig. \ref{widmo}(b)]. For the strong
coupling case of $d=3.25$ the spectrum resembles the one of the
identical rings [cf. Fig. \ref{1eip}(c)].

\begin{figure*}[ht!]
\centerline{\hbox{\epsfysize=50mm
               (a)\epsfbox {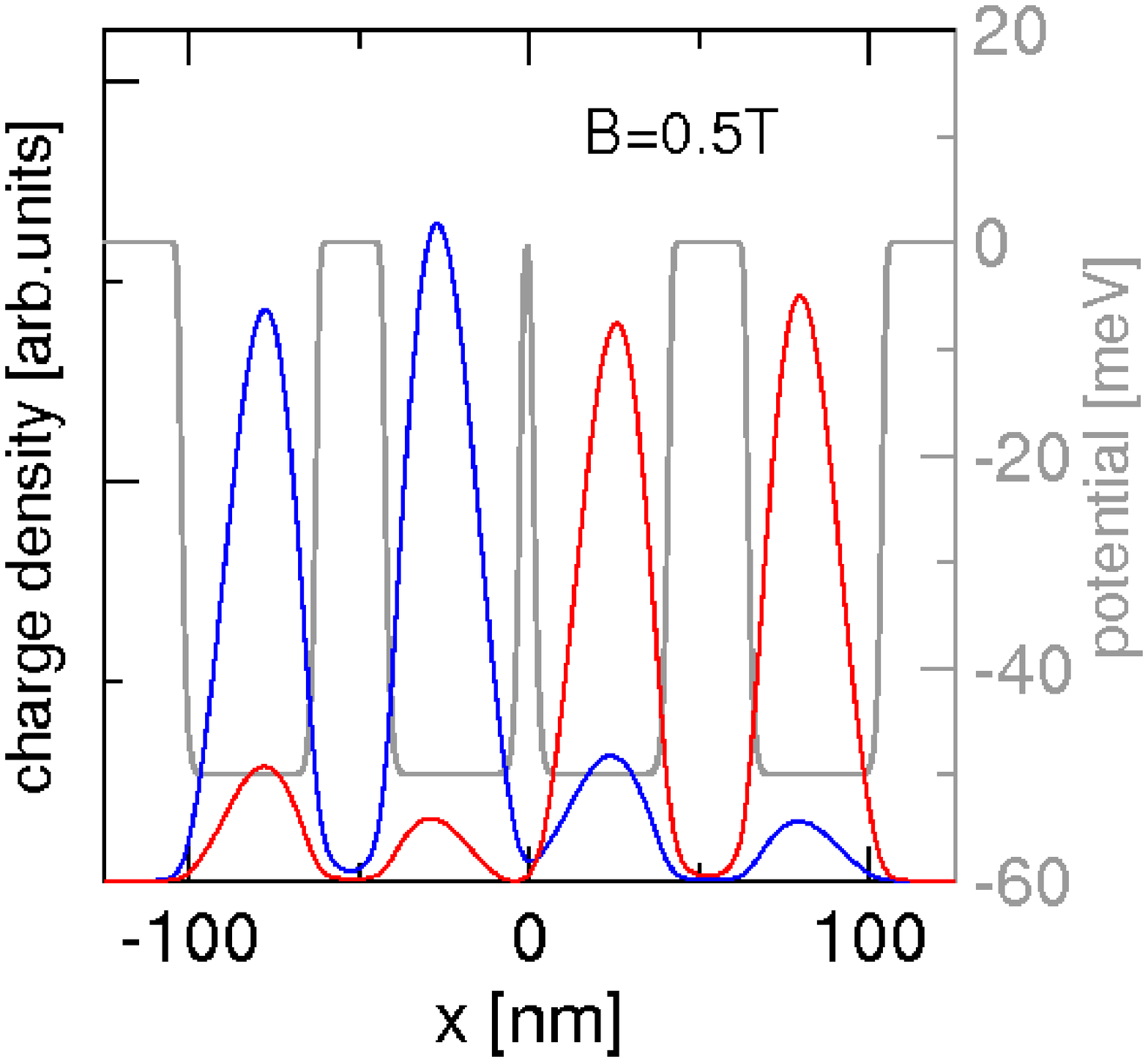}\epsfysize=50mm
               (b)\epsfbox {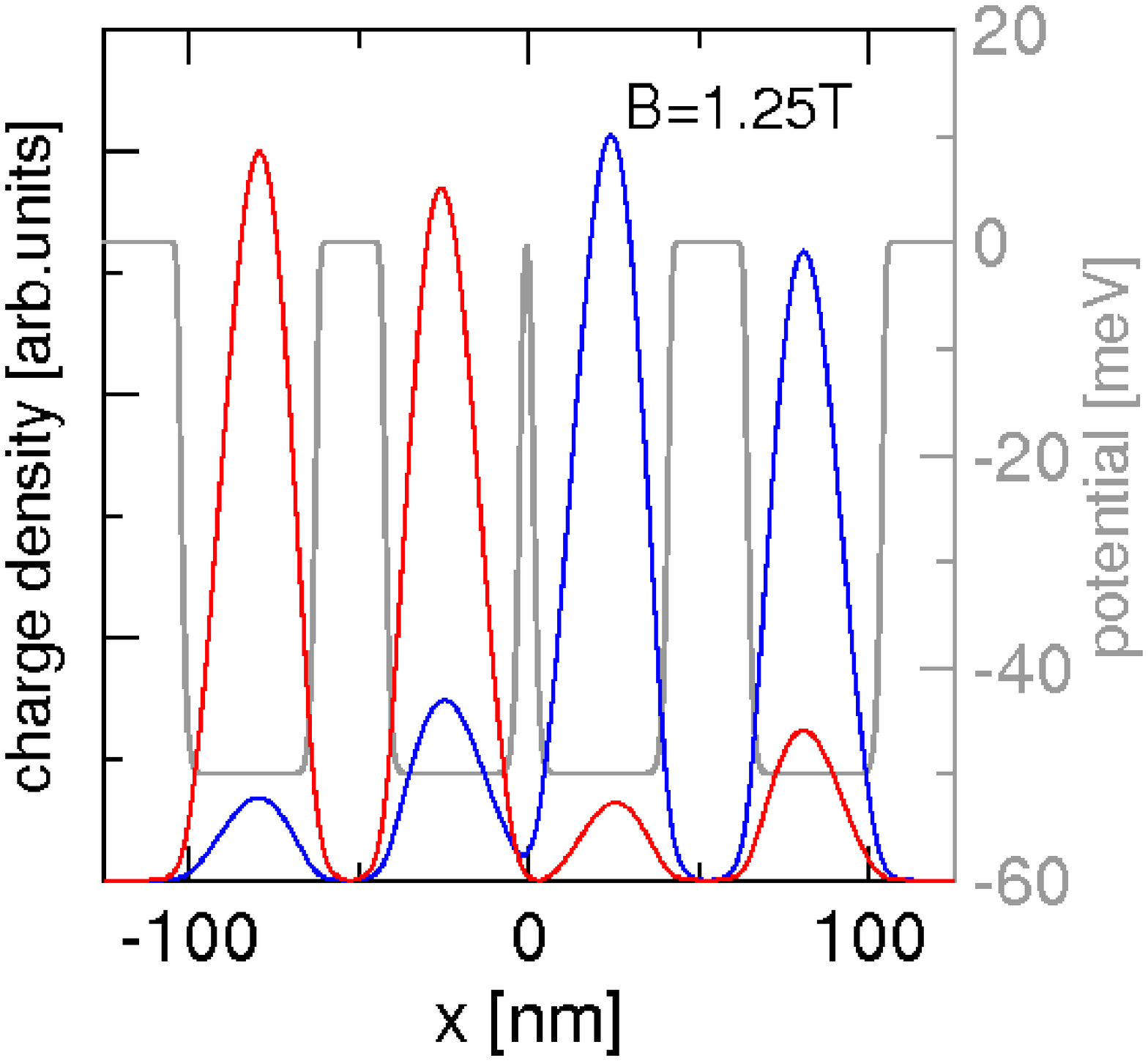}\epsfysize=50mm
               (c)\epsfbox {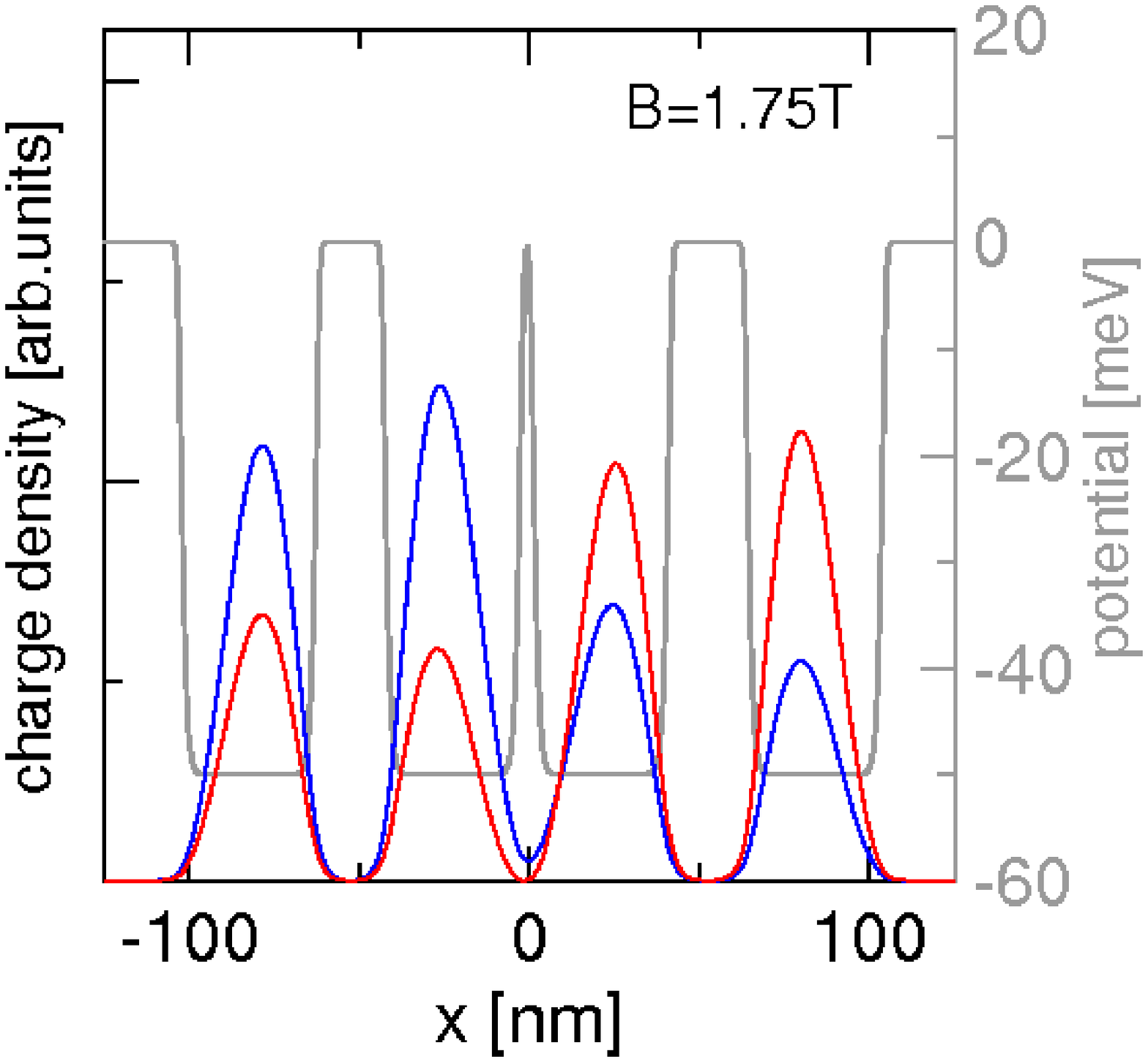}}}
               \centerline{\hbox{\epsfysize=50mm
               (d)\epsfbox {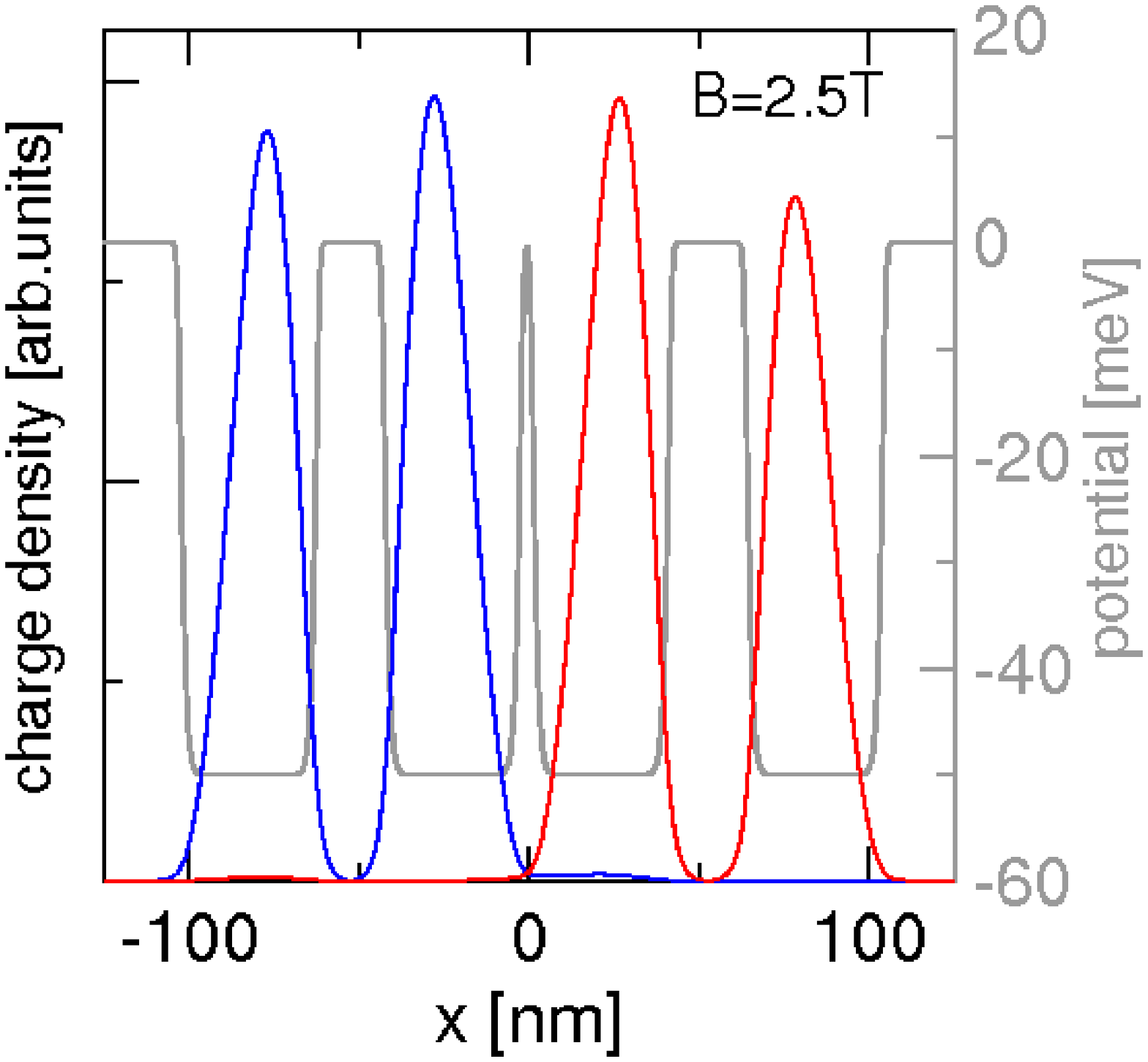}\epsfysize=50mm
               (e)\epsfbox {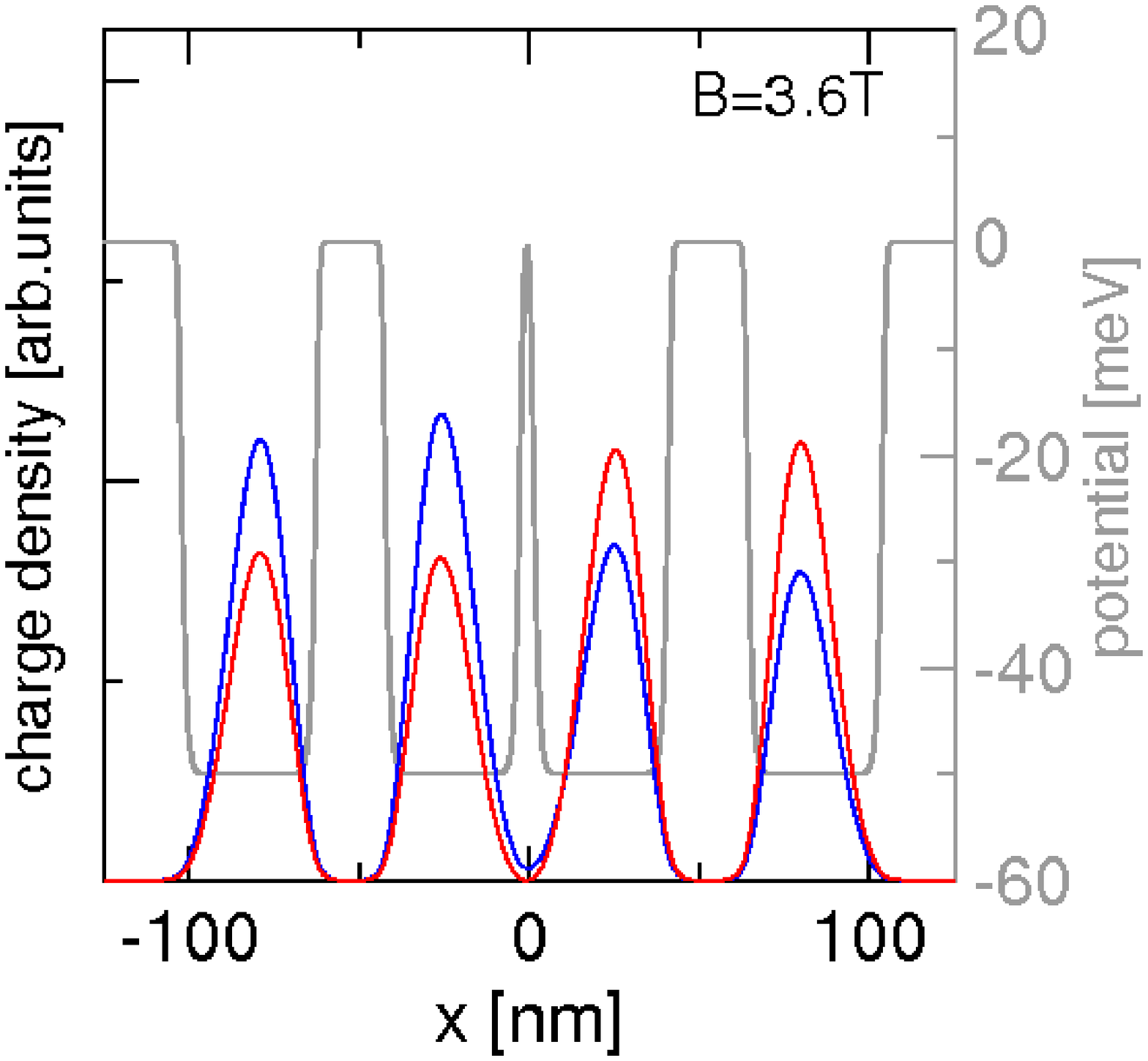}\epsfysize=50mm
               (f)\epsfbox {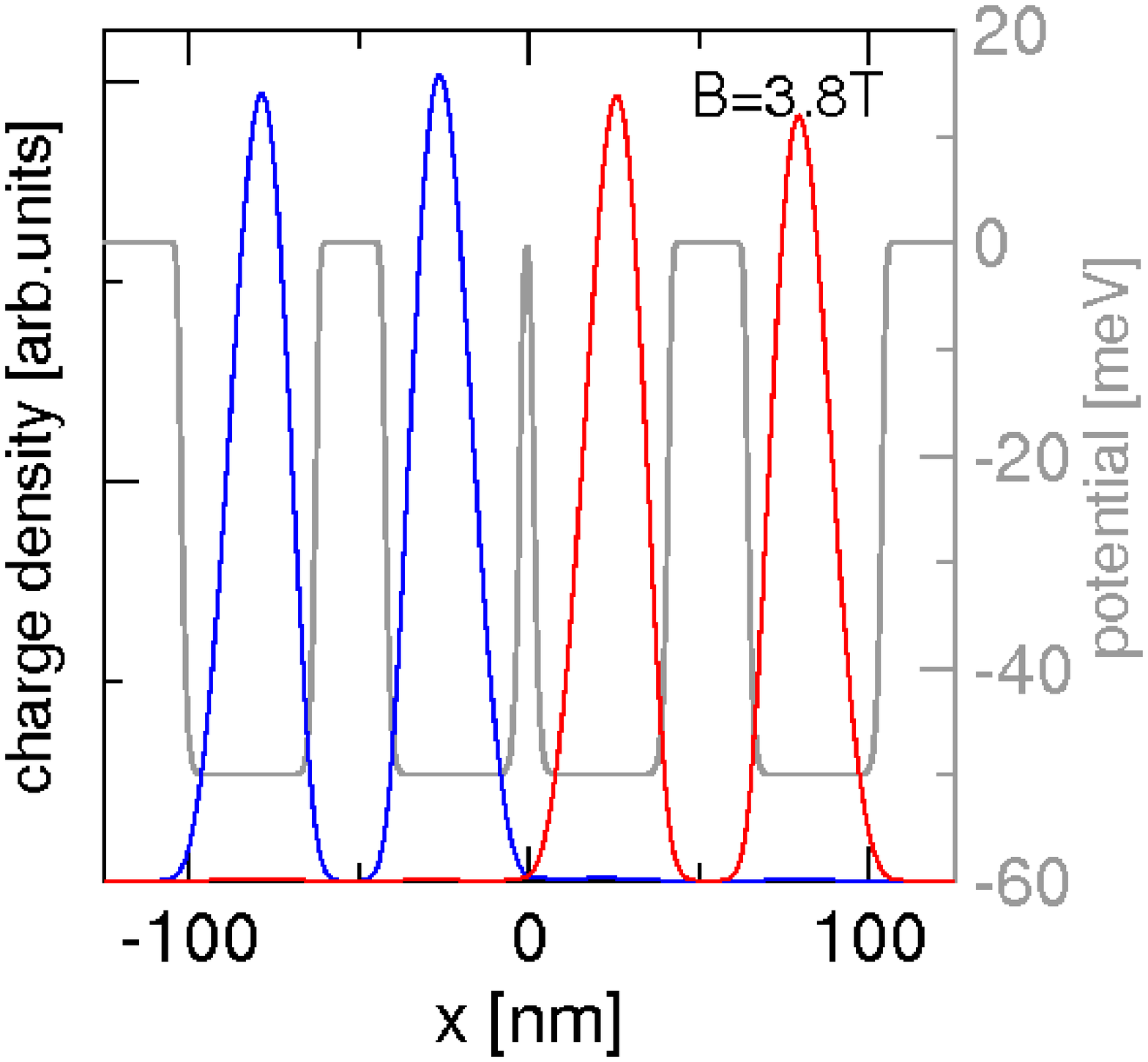}}}
\caption{Cross section of the confinement potential (grey lines) and
the charge density of the single confined electron in the
lowest-energy state (blue curve) and the first excited state (red
curve) for a pair of quantum rings with the radius of the right ring
larger by 5\% and the interring distance parameter $d=3.5$. Panels
correspond to different values of the magnetic field given in the
figure.} \label{1eszift}
\end{figure*}

The shifts of the electron density with the magnetic field between
the rings are illustrated in Fig. \ref{1eszift} for $d=3.5$. For low
magnetic field the electron in both the ground-state and the
first-excited state is present in both the rings [Fig. \ref{1eszift}(a)
for $B=0.5$]. At $B=1$ T we observe an avoided crossing which results
of the change of the order of the energy levels corresponding to
different single-ring angular momenta. For $B<1$ T ($B>1$ T) the
electron in the ground-state occupies preferentially the smaller
(larger) ring [see Fig. \ref{1eszift}(a) and \ref{1eszift}(b)]. As the
magnetic field increases above 1.25 T the ground-state presence of
the electron in the left (smaller) ring is enhanced. Equal electron
distribution is obtained near the center of the avoided crossing for
$B\simeq 1.75$ T [Fig. \ref{1eszift}(c)] and for $B=2.51$ T the
ground-state is totally localized in the smaller ring. Near $B=2.8$
T in [Fig. \ref{widmo}(b)] we see a change of the order of the
energy levels in the ground-state. This is a very narrow avoided
crossing (which in the figure scale appears as a level crossing).
After this avoided crossing the electron in the ground-state
is localized in the larger ring again. Equal distribution of the
ground-state electron between the two rings is found near 3.61 T.
The change of the order of the two energy levels in the ground-state
that occurs here is also accompanied by a narrow avoided crossing. For
the magnetic field increased additionally by only 0.2 T the
interring tunnel coupling is broken and the ground-state and the
first-excited state occupy a single ring. We see that at higher
magnetic fields -- which favor a stronger localization of the
confined states -- the avoided crossing related to electron transfer
 between the rings become narrower both on the magnetic field and energy scales
 (compare Fig. \ref{1eszift}(a-c) and Fig. \ref{1eszift}(d-f)].

\begin{figure*}[ht!]
\centerline{\hbox{\epsfysize=50mm
               \epsfbox {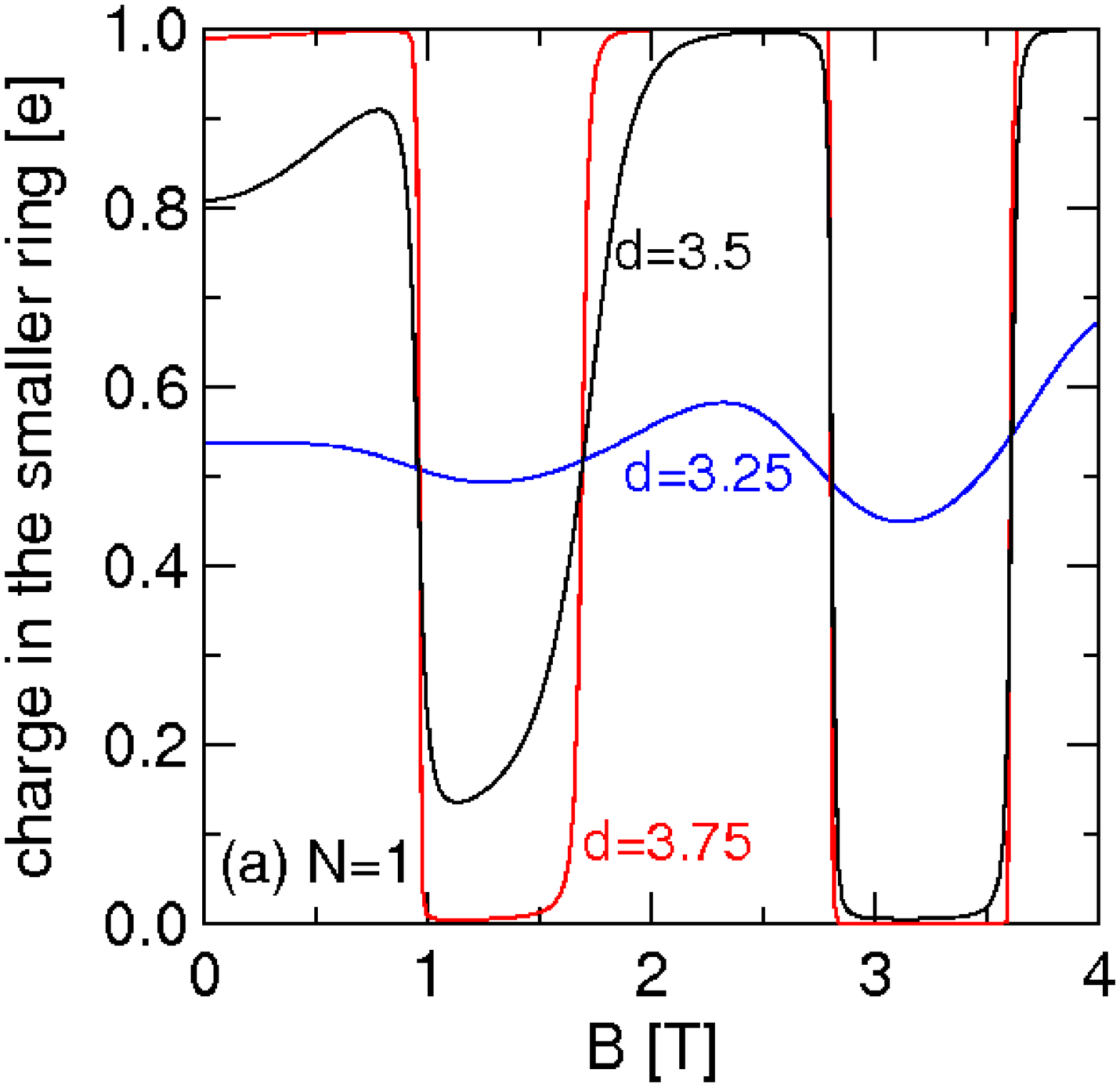}}\hbox{\epsfysize=50mm
               \epsfbox {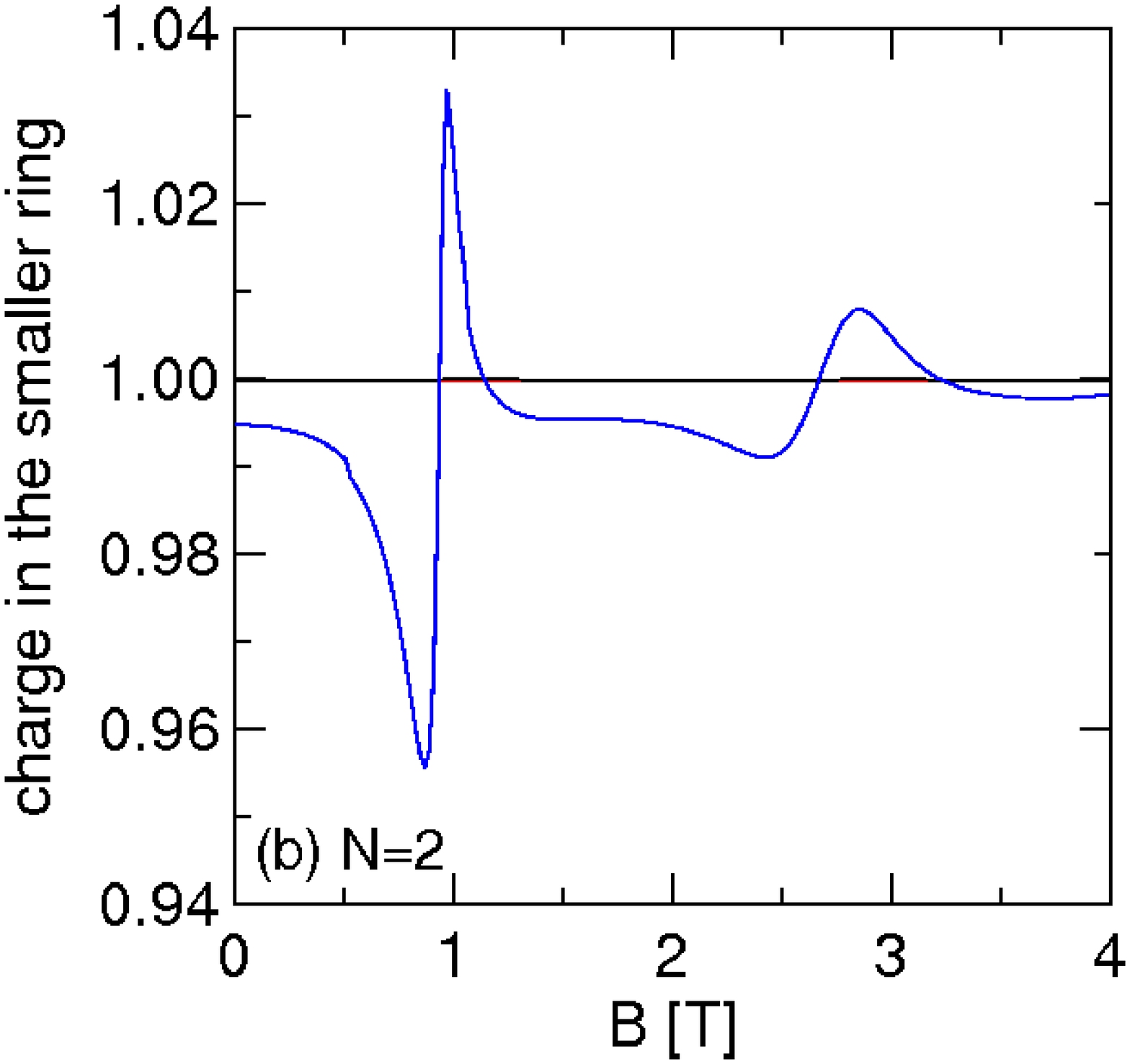}}\hbox{\epsfysize=50mm
               \epsfbox {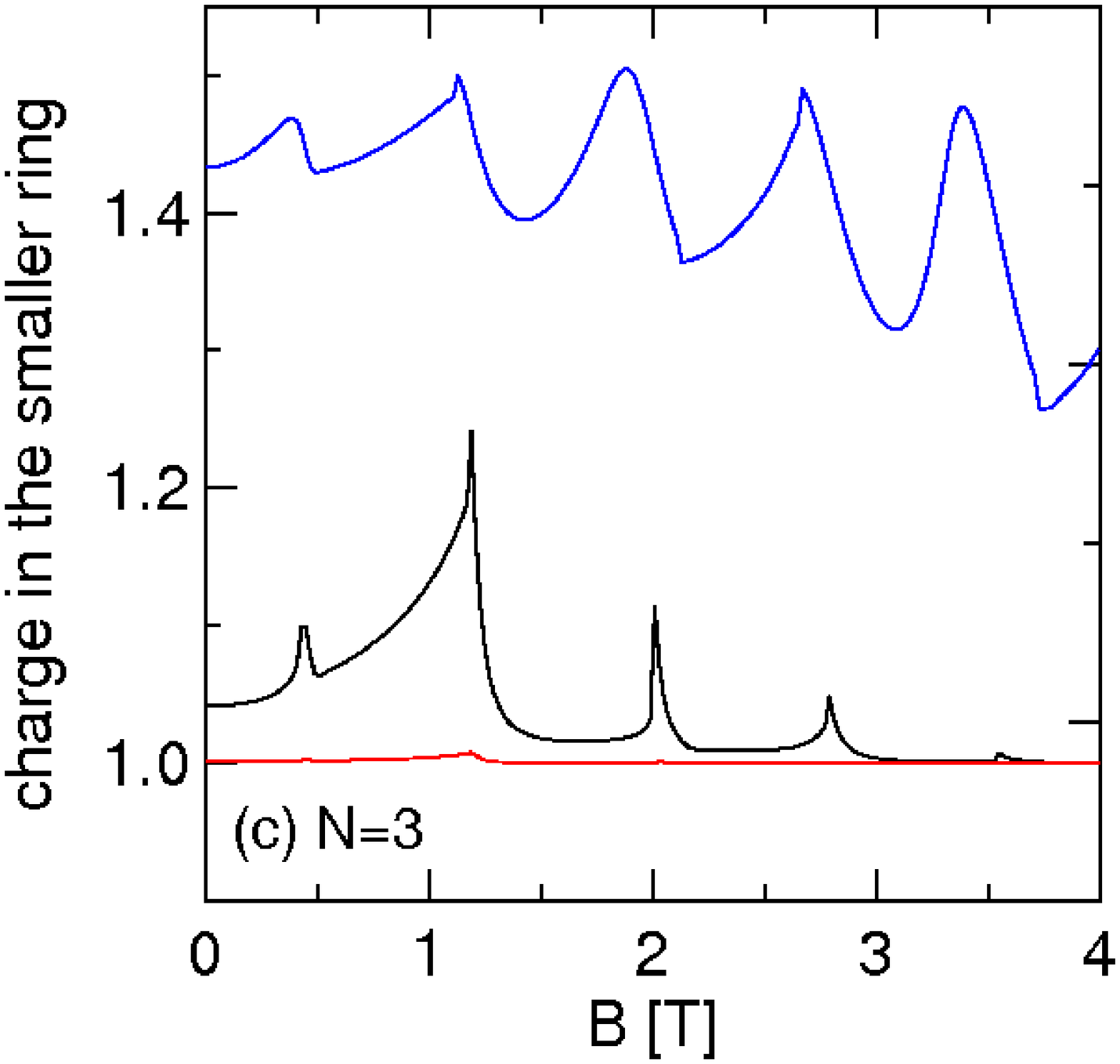}}\hfill}
\caption{Charge localized in the smaller ring for
ground state of the single-electron (a) two-electrons (b) and three-electrons (c) for the ring couple with the radius of
the right ring larger by 5\%. Red, black and blue lines correspond
to $d=3.75$, $d=3.5$ and $d=3.25$, respectively.} \label{leftring}
\end{figure*}

To summarize the shifts of the charge in the single-electron problem
of a single-ring in Fig. \ref{leftring} we plotted the ground-state
charge localized in the left ring for various distance parameters.
For $d=3.75$ the charge in the left ring has a step-like dependence
on $B$ and is either very close to 1 or very close to zero. The
oscillation of the charge for $d=3.5$ becomes less abrupt. For the
strong coupling case $d=3.25$ the charge localized in the left ring
oscillates around $0.5e$. In all the cases the
oscillations of the charge localized in the left ring become more
pronounced as the magnetic field grows, which is related to the
reduction of the tunnel coupling with $B$.

For two-electrons the Coulomb repulsion segregates the electrons between the different
rings. For $d=3.75$ and $d=3.5$ the charge confined in each of the rings is equal to $1e$
[see Fig. \ref{leftring}(b)]. Only for $d=3.25$ deviations off the equal distribution of the rings occur.
We find that these deviations are accompanied by lifting the singlet-triplet degeneracies in the spectrum.
The magnetic field leads eventually to attenuation of the tunnel coupling and segregation
of the carriers between the two dots. 

In the system of three electrons [see Fig. \ref{leftring}(c)] for large interring barrier ($d=3.75$) two-electrons occupy the larger ring
leaving a single electron in the smaller one. For $d=3.5$ the charge accumulated in the smaller ring exceeds
one elementary charge for low magnetic fields.
For $d=3.25$ the electron distribution between the rings becomes closer to $3e/2$ per each ring.
Oscillations of the charge in the smaller ring for $d=3.5$ and $d=3.25$ are associated with the avoided crossings
that appear in the low-spin part of the spectrum as well as with the ground-state spin transitions. The latter produce cusps
on the plotted curves [see Fig. 26(c)].

\section{Summary and Conclusions}

We have studied the system of one, two and three electrons in a planar double ring structure
considering both the tunnel and the electrostatic coupling between the rings using
a Gaussian functions mesh technique.

The presented results indicate that in the system of few electrons there is a
distinct competition of tunnel coupling with 1) the magnetic field which enhances
localization of the occupied states within a single ring 2) the electron-electron
interaction which strengthens the interring barrier and 3) asymmetry effects of the double ring structure which favor electron
localization in one of the rings.

We find that the system undergoes symmetry transitions of the parity and spin in function of the magnetic field,
and that these transitions vary strongly with both $N$ and $d$. At large interring barrier ($d$)
for $N=1$ we find angular momentum transitions for states localized in a single ring. These transitions
are absent in the system of two electrons since their mutual interaction perturbs the rotational symmetry
of each of the rings. $N=2$ ground-state at large $d$ is degenerate with respect to the spin due to
the perfect separation of the electrons.  For $N=3$ at large $d$ the symmetry transformations are present,
and result of the spin transitions within the two-electron subsystem confined in one of the rings.  When the rings are identical at large
$d$ the ground state is degenerate with respect to the parity for any $N$.

In the opposite limit of strong interring coupling and coalesced rings we find lifting of the even-odd degeneracy
for all the studied electron numbers. The symmetry transformations in function of the magnetic field vanish for the single electron, which
occupies the binding orbital and in the limit of coalesced rings -- the quantum dot formed at the contact of the rings.
On the contrary, for two electrons the symmetry transformations occur only when the rings form a single structure.
For three electrons the symmetry transformations are present at any $d$. At large $d$ they are only related to the spin transitions
of the two-electron subsystem and in the strong coupling limit they involve both parity and spin. For $d=3$ the sequence of the
parity and spin ground-state symmetries in function of the magnetic field is identical with the one found for strongly deformed elliptic quantum dots.

We demonstrated that due to the strong dependence of the symmetry transformations on both $N$ and $d$, the confined charge as
well as the distance between the rings should be readily accessible in both single-electron charging and magnetization measurements
performed in function of the magnetic field. In particular the strength of the coupling has an opposite impact on the
chemical potentials of one and two confined electrons. For week tunnel coupling $\mu_1$ has cusps in function of the field
while $\mu_2$ is a smooth function of $B$. In the strong coupling limit the dependence is inverted: chemical potential
of the single electron becomes smooth and the cusps appear on $\mu_2$.

The asymmetry in the depth of the confinement potential favors localization of electrons in the deeper potential well. Less obvious
is the effect of the asymmetry in the size of the rings.
For rings of different radii the magnetic field leads to switching of the ground-state localization from the larger
to the smaller rings. The oscillations of the ground-state localization become more abrupt for stronger magnetic fields
due to the reduction of the tunnel coupling.
Both types of the asymmetry -- in the depth and in the size of the ring -- for large values of $d$ or $B$ -- break the tunnel coupling between the rings,
by tending to localize the electron in one
of the rings -- favoring the atomic type of localization over the molecular (extended) one.
The asymmetry in the double ring potential lifts the degeneracy of the even and odd parity energy levels that is observed at large $d$ for any $N$.
We find that when $d$ is decreased, the asymmetry effects
are reduced. In particular the electron charge between the rings becomes more evenly distributed for the strongly tunnel-coupled asymmetric structures.

\section{Appendix}
In this appendix we explain how the matrix elements for the Coulomb interaction
\begin{equation}
V_{abcd}=\langle \phi_{a}({\bf{r}}_{1}) \phi_{b}({\bf{r}}_{2}) |\frac{1}{r_{12}}|
\phi_{c}({\bf{r}}_{1}) \phi_{d}({\bf{r}}_{2}) \rangle
\label{vabcd},
\end{equation}
are integrated.
In the above expression the single electron wavefunctions are replaced by their linear
combinations (\ref{jedn}):
\begin{equation}
V_{abcd}=\sum_{i,j,k,l=1}^{gN} a_{i}^{*}b_{j}^{*}c_{k}d_{l}
\langle f_{i}({\bf{r}}_{1}) f_{j}({\bf{r}}_{2}) |\frac{1}{r_{12}}|
f_{k}({\bf{r}}_{1}) f_{l}({\bf{r}}_{2}) \rangle
\label{vsuma}
\end{equation}
An interaction integral which appears in (\ref{vsuma}):
\begin{eqnarray}
C_{ijkl}=\langle f_{i}({\bf{r}}_{1}) f_{j}({\bf{r}}_{2}) |\frac{1}{r_{12}}|
f_{k}({\bf{r}}_{1}) f_{l}({\bf{r}}_{2}) \rangle=\\ \nonumber
\int d^{2}{\bf r}_{1}\int d^{2}{\bf r}_{1} f^{*}_{i}({\bf r}_{1}) f^{*}_{j}({\bf r}_{2})
\frac{1}{r_{12}}f_{k}({\bf r}_{1}) f_{l}({\bf r}_{2})
\label{cijkl}
\end{eqnarray}
can be calculated in following way.
First, we should substitute if place of  $f_{i}$, $f_{j}$, $f_{k}$, $f_{l}$ and $\frac{1}{r_{12}}$
their invert Fourier transform. Next step is to integrate over  ${\bf r}_{1}$ and  ${\bf r}_{2}$
variables.  After these we get the formula on $C_{ijkl}$:
\begin{equation}
C_{ijkl}=B \int d^{2}{\bf{k}}\frac{1}{k}\exp\bigg(-\frac{\sigma^{2}k^{2}}{2} \bigg)
\exp\bigg(i{\bf{k}}\cdot{\bf{R}} \bigg)\exp\bigg(\frac{\gamma \sigma^2}{2} {\bf{k}}\cdot{\bf{r}}
\bigg)
\label{ck1}
\end{equation}

where
\begin{eqnarray}
{\bf{R}}&=&(X_{ik}-X_{jl},Y_{ik}-Y_{jl}) \nonumber \\
              &=&R(\cos\beta,\sin\beta)\\
{\bf{r}}&=&(y_{jl}-y_{ik},x_{ik}-x_{jl}) \nonumber \\
             &=&r(\cos\alpha,\sin\alpha)\\
{\bf{r}}_{ik}&=&{\bf{R}}_{i}-{\bf{R}}_{k}\\
{\bf{r}}_{jl}&=&{\bf{R}}_{j}-{\bf{R}}_{l}\\
{\bf{R}}_{ik}&=&\frac{{\bf{R}}_{i}+{\bf{R}}_{k}}{2}\\
{\bf{R}}_{jl}&=&\frac{{\bf{R}}_{j}+{\bf{R}}_{l}}{2}\\
\gamma&=&-\frac{eBa_{B}^{2}}{2\hbar}
\end{eqnarray}
$a_{B}$ is Bohr radius,\\
and
\begin{eqnarray}
B & = & \frac{\pi\sigma^{4}}{2}\exp\bigg(\frac{-\gamma^{2}\sigma^{2}}{4}({\bf{r}}_{ik}^{2}+{\bf{r}
} _ {jl} ^ { 2 }) \bigg) \nonumber \\
&&\times \exp\bigg(i\gamma (-Y_{ik}x_{ik}-Y_{jl}x_{jl}+X_{ik}y_{ik}+X_{jl}y_{jl})  \bigg) \nonumber
\\
&& \times \exp\bigg( -\frac{r_{ik}^{2}+r_{jl}^{2}}{4\sigma^2} \bigg).
\end{eqnarray}
The expression {\ref{ck1}} can be further simplified by doing transformation the integral form
Cartesian to the cylindrical coordinates and then by integrating over the $k$ variable:
\begin{equation}
C_{ijkl}=B\frac{\sqrt{2\pi}}{2\sigma}\int_{0}^{2\pi}d\varphi
\exp\bigg(\frac{A^{2}(\varphi)}{2\sigma^2} \bigg)\bigg[1-\mathrm{erf}\bigg(\frac{A(\varphi)}{\sqrt{2}\sigma}
\bigg) \bigg]
\label{cost}
\end{equation}
The function $A(\varphi)$ is defined as below:
\begin{equation}
A(\varphi)=-i R\cos(\varphi-\beta)-\frac{\gamma\sigma^2}{2}r\cos(\varphi-\alpha)
\end{equation}
where  $\gamma=-\frac{eB}{2\hbar}$.
The value of integral  (\ref{cost}) is calculated numerically.


\begin{thebibliography}{00}
\bibitem{opis} $(1+P_{12})$ in Eq. (\ref{twv}) ensures the symmetry of
the wave function with respect to the electron exchange and
simultaneously even spatial parity of the state. For $d>3.5$ in the
ground state the even singlet is degenerate with the odd triplet.
The odd triplet is obtained for $(1+P_{12})$  replaced by
$(1-P_{12})$. As long as the electron tunneling between the rings is
negligible (as long as the radial functions $\chi_l$ and $\chi_r$
are separated) the two-electron density is the same for both the
states.
\end{thebibliography}
\end{document}